\documentclass[fleqn,usenatbib,useAMS]{mnras}

\usepackage{bm}
\usepackage{hyperref}
\usepackage{multicol}
\usepackage{pdflscape}
\usepackage{orcidlink}
\usepackage{newtxtext,newtxmath}

\usepackage[T1]{fontenc}
\usepackage{ae,aecompl}

\newcommand{\pI}{{\hyperlink{ac:pI}{Paper I}}}

\newcommand{\dd}{{\rm d}}
\newcommand{\avg}[1]{\ensuremath{\mathinner{\left\langle #1 \right\rangle}}}
\newcommand{\paren}[1]{\ensuremath{\mathinner{\left( #1 \right)}}}

\newcommand{\vrms}{v_{\rm rms}}
\newcommand{\nks}{N_{\rm kicks}}
\newcommand{\dv}{|{\rm \Delta \mathbf{v}}_i|}

\renewcommand{\eqref}[1]{Eq.~\ref{#1}}

\title[Observational Constraints for White-Dwarf Natal Kicks]{White Dwarf Natal Kicks as Velocity-Space Random Walks. II. \\ Binary Stellar Evolution and Observational Kick Constraints}

\author[Pham \& Majeski et al.]{Dang Pham$^{1, 2}$
\thanks{E-mail: \href{mailto:dang.pham@colorado.edu}{dang.pham@colorado.edu}}
\orcidlink{0000-0002-0924-8403},
Stephen Majeski$^{2}$
\thanks{E-mail: \href{mailto:stephen.majeski@colorado.edu}{stephen.majeski@colorado.edu}}
\orcidlink{0000-0002-7879-060X},
Taeho Ryu$^{2, 1}$
\orcidlink{0000-0003-2012-5217},
Gabriel Tomassini$^{3}$
\orcidlink{0009-0003-6995-1840},
Andrea Chiavassa$^{3}$
\orcidlink{0000-0003-3891-7554}
\\
$^{1}$Department of Astrophysical and Planetary Sciences, University of Colorado, 391 UCB, Boulder, CO 80309-0391, USA\\
$^{2}$JILA, University of Colorado and National Institute of Standards and Technology, 440 UCB, Boulder, CO 80309-0440, USA\\
$^{3}$Université Côte d’Azur, Observatoire de la Côte d’Azur, CNRS, Lagrange, CS 34229, Nice, France
}

\pubyear{2026}
\begin{document}
\label{firstpage}
\pagerange{\pageref{firstpage}--\pageref{lastpage}}
\maketitle

\begin{abstract}

Wide binaries containing white dwarfs (WDs) provide a pathway to exploring post-main-sequence (MS) natal kicks associated with asymmetric mass loss.
In this work, we develop a binary population synthesis model that involves episodic mass loss and velocity kicks during the thermally-pulsing asymptotic-giant-branch (TP-AGB) phase of stellar evolution, when mass loss is expected to be most significant.
We compare our synthesized populations against the \textit{Gaia} DR2 and DR3 data for wide WD-MS and WD-WD binaries, varying the number, strength, and directionality of TP-AGB kicks.
We find that the observed separation distribution can constrain the kick strength, but it alone is insufficient to also constrain the kick number and dimensionality.
We therefore additionally constrain our models using the observed eccentricity distributions.
Models where mass loss induces a random walk in the star's velocity are most consistent with observed eccentricity constraints for WD-MS binaries.
No models are found to agree with the WD-WD eccentricity constraint.
Our results suggest that at least 10 randomly directed kicks take place during the TP-AGB phase, leading to a cumulative root-mean-square kick velocity of approximately $0.7$ km/s. 
These findings constrain the nature of WD natal kicks, connecting the orbital distributions of evolved wide binaries to asymmetric mass loss during the final stages of stellar evolution.
\end{abstract}

\begin{keywords}
stars: kinematics and dynamics --stars: evolution --
stars: mass-loss --
stars: binaries --
stars: white dwarfs -- 
methods: statistical
\end{keywords}

\section{Introduction}\label{sec:intro}

In order for a main-sequence (MS), low-to-intermediate mass star to become a white dwarf (WD), it must first pass through the asymptotic giant-branch (AGB) phase of stellar evolution. The AGB phase is marked by turmoil: the star loses an order-unity fraction of its mass through a combination of stellar winds and episodic mass-ejection events, all within a fraction of its main-sequence lifetime~\citep{Hofner2018,Decin2019,Decin2021,Freytag2023}. The late, thermally pulsing (TP) stage of AGB evolution is particularly violent: observations indicate a sharp increase in the rate of mass loss during this period through the episodic ejection of shells of stellar material~\citep[e.g.,][]{maercker2012,maercker2024}.

Given the relatively short lifetime of the TP-AGB phase ($\tau_{\rm AGB}\sim10^5{-}10^{6}$ yrs or $\lesssim 10^{-4}\tau_{\rm MS}$) and the complex physics involved in modeling it, it is difficult to study directly, both via observation and theory. That said, TP-AGB evolution can be studied indirectly, because mass loss leads to observable dynamical changes in systems that once contained AGB stars. When asymmetry is present in the ejected stellar material, it causes the TP-AGB star to recoil in the opposite direction, in what is known as a WD ``natal kick''. Examples of such asymmetry include clumpy, irregular molecular line emission from the winds of AGB stars~\citep{Ohnaka2016,Ohnaka2017,Ohnaka2025}, and partial detached shells of stellar material around TP-AGB stars, presumably resulting from thermal pulses~\citep{maercker2024}. The resultant natal kicks have been shown to modify the radial distributions of WDs in globular clusters~\citep{Heyl2007} and even eject them from open clusters~\citep{fellhauer2003}. By comparing analytical and numerical models of WD natal kicks to observed WD populations, properties of the kicks can be indirectly determined, such as their strength and the period of time over which they occur~\citep{Davis2008,Fregeau2009}.

Recently, wide stellar binaries containing WDs have been shown to be a particularly effective probe of TP-AGB stellar evolution and the associated natal kicks. \citet{ElBadry2018} invoked adiabatic mass loss, coupled with a single, impulsive natal kick of roughly 0.9 km/s, to explain the separation distributions of wide WD--MS and WD--WD binaries detected by \textit{Gaia}. This approach was refined further by \citet{hwang2025}, where the authors instead modeled the natal kick as a slow, unidirectional acceleration that occurs over the TP-AGB lifetime, and further compared the resultant model to the eccentricity distribution of observed WD binaries. It was once again found that natal kicks in the range of $0.25-4$ km/s were necessary to explain WD wide binary separation distributions. In addition, they also determined that for separations beyond ${\sim}10^3$ au, WD binaries appear to converge toward a thermal eccentricity distribution as a result of these post-MS stellar-evolution processes.

In our preceding companion paper (\citealt{paper1}; hereafter \hypertarget{ac:pI}{Paper I}), we considered a different kind of natal kick altogether. We assumed that many, impulsive, randomly directed kicks occur over the TP-AGB lifetime as a result of an episodic mass-loss process. By virtue of the assumed kicks being many and random, we were able to develop an analytic Fokker-Planck theory for the evolution of distributions of bodies orbiting WD progenitors. From this theory, we recovered key features of the separation distribution of WD binaries described in previous works, namely, that a broken power-law with a $-2$ tail is produced in the distribution of WD wide-binary separations~\citep{ElBadry2018}, and that the distribution of WD binary eccentricities appears to become thermalized beyond ${\sim}10^3$ au~\citep{hwang2025}. We additionally probed how random-walking WD natal kicks might affect the orbital distributions of planets and debris populations in our own solar system and exoplanetary systems.

Although the Fokker-Planck approach of \pI~allowed us to probe the dynamical physics of random-walking WD natal kicks in varied systems, it is not well-suited for direct comparison with binary observations. This is because each Fokker-Planck solution is obtained only for binaries of a particular mass, mass ratio, and companion type (WD or MS). In contrast, observed binary distributions contain contributions from a considerable range of values in all of the aforementioned parameters. Thus, for a quantitative comparison to observed distributions, it is more reasonable to work directly with a simulated sample of such binaries that explores those parameters thoroughly. 

To that end, in this paper (Paper II), we develop and implement a binary population synthesis (BPS) model that includes varying numbers of randomly directed kicks and accompanying mass loss during the TP-AGB phase in order to directly test the random-walk paradigm of WD natal kicks against observed binaries. We compare the synthesized WD wide-binary populations to those observed by \textit{Gaia} to constrain both the strength of the individual kicks as well as how many occur during the TP-AGB lifetime.

The remainder of this paper is structured as follows. In \S\ref{sec:bps}, we introduce the binary population synthesis model used to reproduce the observed binary orbital-parameter distributions, given our random-walk WD natal kick framework. In \S\ref{sec:fits}, we fit the resultant model populations to the observed \textit{Gaia} WD binary separation distributions in order to determine, for a given number of random kicks, the best-fit total (root-mean-square) kick strength. With these best-fitting kick strengths, we compare the predicted eccentricity distributions with the observational constraints to assess which kick model and number of kicks are favored in \S\ref{sec:ecc}. In \S\ref{sec:cedis}, we report the fraction of simulated systems that can enter common envelope evolution (CE) and experience binary dissolution.
In \S\ref{sec:disc}, we compare our results with \pI, address the limitations of our model, and summarize this work.

\section{Binary Population Synthesis}\label{sec:bps}

In this section, we describe the binary population synthesis model used to simulate observed WD--MS and WD--WD binaries.
First, the MS--MS initial conditions are introduced.
Then, we present the post-MS mass-loss and velocity-kick prescriptions implemented.
Finally, we discuss the orbital evolution methodology, boundary conditions, and stopping conditions.

\subsection{Initial Conditions}
In our model, we use the same initial stellar-population prescription as that of \citet{hwang2025}, which builds upon that of \citet{ElBadry2018}.
We summarize this five-step procedure below.

First, initial binary masses are randomly selected from the initial-mass function by \citet{Kroupa2001}, in the range $0.3 - 7.2 M_\odot$ (i.e., MS stars that can become WDs).
The stellar masses in a binary are drawn independently, motivated by observational evidence from \citet{Moe2017} showing that masses are largely uncorrelated in wide binaries.

Second, in systems where the primary mass is greater than $0.75 M_\odot$, orbital periods (equivalently, semi-major axes) are drawn from a log-normal distribution described by $\avg{\log(P/\mathrm{day})} = 4.8$ and $\sigma = 2.3$ as found by \citet{Duquennoy1991}.
Here, $\log$ is the base-10 logarithm, and $\avg{\cdot}$ and $\sigma$ are the mean and standard deviation of the log-normal distribution, respectively.
In systems where the primary star's mass is lower than $0.75 M_\odot$, the binary's orbital period is drawn from a log-normal distribution with $\avg{\log(P/\mathrm{day})} = 4.1$ and $\sigma = 1.3$ \citep{Fischer1992}.

Third, orbital eccentricity is drawn for each binary according to \citet{Hwang2022}.
Specifically, for a MS--MS binary with a given semi-major axis, we draw its eccentricity from a distribution with probability density $P(e)\propto e^\alpha$.
The eccentricity power-law index $\alpha$ is technically a function of projected separation \citep{Hwang2022}.
Following \S3.2 in \citet{hwang2025}, however, we approximate the separation dependence of $\alpha$ as a dependence on semi-major axis.
Elsewhere in our models, we do not neglect the difference between semi-major axis and separation.

Fourth, the remaining orbital elements are drawn isotropically.
That is, the argument of pericenter, longitude of ascending node, and mean anomaly are drawn uniformly in the range $[0, 2\pi]$.
We draw the cosine of the orbital inclination uniformly over $[-1,1]$ to sample isotropic orbital angular-momentum directions.

Fifth, a binary's birthday is drawn uniformly over the last 12 Gyr, which is approximately the age of the Galaxy.
This corresponds to a constant star formation rate over the same interval, as noted by \citet{ElBadry2018}.
For each star, we calculate the MS lifetime using the prescription of \citet{Lamers2017}.
If neither star in a binary completes its MS evolution within 12 Gyr, we discard the system and draw a replacement.
This procedure effectively introduces a lower stellar mass limit at $\approx 0.9 M_\odot$, as not all stars would have begun post-MS evolution by the present day.
This procedure yields a sample of $10^8$ stellar binaries that can reach the WD--MS or WD--WD stage within the Galaxy's age.
We then apply post-main-sequence (post-MS) mass loss and kicks to this sample.

\subsection{Mass Loss Model}\label{sec:massloss}
At the end of a star's MS lifetime, we set its mass to the value it should take at the beginning of the TP-AGB phase using MIST, a grid of MESA stellar evolution models \citep{Paxton2011, dotter2016}.\footnote{The dependence of post-MS mass loss on initial stellar mass remains an open research topic, with ongoing approaches employing simulations \citep[e.g.,][]{Marigo2007, Cui2026} and observational constraints on the initial-final mass relation \citep[e.g.,][]{Kalirai2014, Miller2026a, Miller2026b}. We adopt MIST for its consistent stellar masses and evolutionary timescales within the MESA framework.}
This accounts for the loss of a few percent of its initial mass that occurs during the main sequence.
Motivated by \pI~and stellar evolution models, this pre-TP-AGB mass loss is assumed to occur slowly and thus, isotropically, only causing adiabatic expansion of the binary's semi-major axis.
Adiabatic mass loss increases the semi-major axis, $a$, while conserving $(M_1 + M_2)a$, where $M_1+M_2$ is the total binary mass.
After applying this initial mass loss, the star is now at the beginning of its TP-AGB phase.

In each realization of a star's TP-AGB phase, we apply $N_{\rm kicks}$ episodic, individual mass-loss events and velocity kicks.
For a given \textit{total} TP-AGB mass loss obtained from MIST, $\Delta M_{\rm TP-AGB}$, we assume that each individual event features the same amount of mass lost, 
\begin{equation}
    \Delta M = \frac{\Delta M_\mathrm{TP-AGB}}{N_\mathrm{kicks}}.
\end{equation}
Note that, because this mass loss is assumed to occur here in impulsive events, it is distinct from the smooth, steady, wind-driven mass loss assumed of some previous works~\citep[e.g.,][]{hwang2025,oconnor2026}.

\subsection{Velocity Kick Models}\label{sec:kickmodels}

During each of the aforementioned mass-loss events, we apply velocity kicks to the star.
When there is a single, instantaneous kick, $N_\mathrm{kicks}=1$, we follow \citet{ElBadry2018} and draw the kick magnitude $|\Delta \mathbf{v}|$ from a Maxwellian distribution
\begin{equation}
    |\Delta \mathbf{v}| \sim \mathrm{Maxwell}(v_\mathrm{rms})
\end{equation}
where $\sim$ denotes that the magnitude of each kick, $|\Delta \mathbf{v}|$, is drawn from a Maxwellian distribution described by the parameter $v_\mathrm{rms}$ with a probability distribution function
\begin{equation}
    P_\mathrm{Maxwell}(|\Delta \mathbf{v}|; v_\mathrm{rms}) = \sqrt{\frac{2}{\pi}} \frac{|\Delta \mathbf{v}|^2}{(v_\mathrm{rms}/3)^3} \exp\paren{\frac{-|\Delta \mathbf{v}|^2}{2 (v_\mathrm{rms}/3)^2}}.
\end{equation}

Given $|\Delta \mathbf{v}|$, the direction of the kick is drawn randomly and isotropically in three dimensions.
Note that \cite{ElBadry2018, oconnor2026} used the dispersion parameter $\sigma=v_\mathrm{rms}/\sqrt{3}$ to describe a Maxwellian.
As $\sigma$ can be confused for the standard deviation, we use $v_\mathrm{rms}$. 

For multiple kicks ($N_\mathrm{kicks}>1$), we consider three different scenarios corresponding to the kick dimensionality, $d$. For each choice of $d$, we will assume that all kicks to a particular star are of the same magnitude $|\Delta \mathbf{v}|$.

In one dimension ($d=1$), we assume that all kicks are applied in the same direction, which we denote as ``unidirectional''.
The direction of the kicks is chosen isotropically in three dimensions, but all kicks are applied in that same direction over the TP-AGB lifetime.
The total kick magnitude imparted on a star is drawn from a Maxwellian
\begin{equation}
    v_\mathrm{total, \mathrm{1d}} = \Bigg|\sum_{i=1}^{N_\mathrm{kicks}} \Delta \mathbf{v}_i\Bigg| \sim \mathrm{Maxwell}(v_\mathrm{rms}).
\end{equation}
Therefore, each kick has a magnitude
\begin{equation}\label{eq:dv1d}
    |\Delta \mathbf{v}_i| = v_{\mathrm{total},1d} / N_\mathrm{kicks}.
\end{equation}
We emphasize that this scenario is \textit{not} a random walk (a one-dimensional random walk would require kicks in opposing directions).
We also note that this model is different from the previous continuous white dwarf recoil model in the current literature \citep[][]{hwang2025, oconnor2026}.
First, this paper treats kicks and mass loss as discrete events, whereas previous works assumed a continuous event occurring over the entire (TP--)AGB lifetime.
Second, our mass loss and velocity kicks are independent, whereas previous papers couple continuous mass loss to the recoil kick (see Eq. 4 in \citealt{hwang2025} and Eq. 15 in \citealt{oconnor2026}).

When there are many kicks in two dimensions, $d=2$, we first choose a random plane in which the kicks occur.
The normal to the plane is drawn randomly and isotropically in three dimensions.
Then, all kicks are applied in random directions that are restricted to this plane. This case therefore \textit{does} correspond to a random walk, albeit one constrained to two dimensions.
This case represents a scenario where mass-loss events are preferentially orthogonal to the rotational axis of the star, motivated by an analogy with the Sun's coronal mass ejections and their tendency to occur most often at low latitudes~\citep{hundhausen1993,stcyr1999,AbdelSattar2018}.

In the three-dimensional case, $d=3$, we draw the direction of each kick independently and isotropically.
This represents random kicks with uncorrelated asymmetries in each episodic mass-loss event, as first proposed by \citet{hwang2025}.

For the $d=2$ and $d=3$ random walks, independent kick directions imply $v_\mathrm{rms}^2=N_\mathrm{kicks}|\Delta\mathbf{v}_i|^2$.
The relationship between the per-kick magnitude and the cumulative root-mean-square velocity is therefore distinct from the $d=1$ unidirectional kick:
\begin{equation}\label{eq:dvrw}
    |\Delta \mathbf{v}_i| = v_\mathrm{rms} / \sqrt{N_\mathrm{kicks}}.
\end{equation}
Thus, each kick model can be parameterized by either the per-kick magnitude $|\Delta\mathbf{v}_i|$ or the cumulative root-mean-square kick velocity, $v_\mathrm{rms}=\sqrt{\avg{v_\mathrm{total}^2}}$, where the average is taken over the simulated stellar population.

Finally, we also implement the adiabatic mass-loss model with a single kick in \citet{ElBadry2018}, henceforth referred to as the ER18 model.

\subsection{Orbital Evolution}

After each mass-loss and velocity-kick event, we recalculate the binary's orbital elements and continue evolving the orbit until the next event.
For a bound binary, we advance the mean anomaly to the time of the next event.
We then calculate the positions of both stars, with orbital elements calculated using the $N$-body code REBOUND \citep{Rein2012}.
When the binary is unbound, the orbit is evolved with episodic mass loss and kicks with the adaptive time-step IAS15 integrator \citep{Rein2015, Pham2024a}.

Our orbital evolution method differs from those used in previous work and offers several advantages.
\citet{ElBadry2018} adapted the Maxwellian supernova-kick prescription of \citet{Hurley2002} to model a single impulsive WD kick, using a statistical conversion to obtain projected separations (see their Appendix~B).
Our model tracks orbital evolution for multiple-kick models and calculates intrinsic and projected separations directly.
In addition, this model naturally recovers the slow recoil model in the high-$\nks$ limit, as previously mentioned.

Our simulation has two stopping conditions.
First, we continue evolving unbound binaries while their separation remains below $10^6$ au, because subsequent kicks may rebind them.
Although the threshold of $10^6$ au is arbitrary, it is one order of magnitude larger than the typical binary dissolution scale associated with stellar flybys \citep[cf.][]{modak2023}.
We therefore consider an unbound binary separated by $>10^6$ au unlikely to become bound again.
Second, we remove a binary if its separation or pericenter distance falls below $1$ au during the post-MS evolution, because it may undergo common envelope (CE) evolution \citep{Hofner2018}.
As CE can be complex with additional physics governing the orbital evolution and survival of the binaries \citep{Ivanova2013, Grondin2024, Grondin2026}, we are unable to properly simulate binaries experiencing this process.

We neglect tidal evolution, assuming it does not affect our main results for wide binaries, although it can be important for close binaries \citep{Hut1981, Verbunt1995, Vick2020, oconnor2026}.
We also neglect stellar flybys and the Galactic tide \citep{Heisler1986, modak2023} during TP-AGB evolution because their characteristic timescales exceed the phase's $10^5$--$10^6$ yr lifetime.
In the Solar neighborhood, flybys within $10^4$ au occur roughly every $10^7$ yr \citep{Zink2020, Brown2022}, while the Galactic tide acts on binaries at comparable separations over $\sim10^8$ yr \citep{Heisler1986}.

\section{Fitting the Separation Distribution}\label{sec:fits}

\subsection{Fitting to Observation}

In this section, with $\nks$ and $\vrms$ as parameters, we fit the binary population synthesis model described in \S\ref{sec:bps} to the projected separation distributions of WD--MS and WD--WD binaries in the \textit{Gaia} DR2 catalog of \citet{ElBadry2018}.
We restrict this sample to projected separations between $10^{1.5} - 10^{4.5}$ au.
Furthermore, although a more recent catalog of wide binaries containing WDs is available \citep{ElBadry2021}, we use the catalog of \citet{ElBadry2018} for our separation analysis to allow a direct comparison with their WD natal-kick constraints.

Binaries in this catalog must be spatially resolved and satisfy the quality cuts imposed by \citet{ElBadry2018}.
The catalog is therefore incomplete, with companion detectability depending on angular separation and magnitude contrast in \textit{Gaia} observations.
Those authors developed a methodology to characterize and account for this incompleteness, which we employ in our fitting procedure.
We then fit the observational data via maximum-likelihood to find the best-fitting $\vrms$ given a kick model (specified by $\nks$ and $d$).
Knowing the catalog's incompleteness, we also perform mock observations of our binary population synthesis model to directly compare to data.
We describe this statistical procedure in detail in Appendix \ref{appendix:sep}. 

\subsection{Results}

\begin{figure}
    \centering
    \includegraphics[width=1.\linewidth]{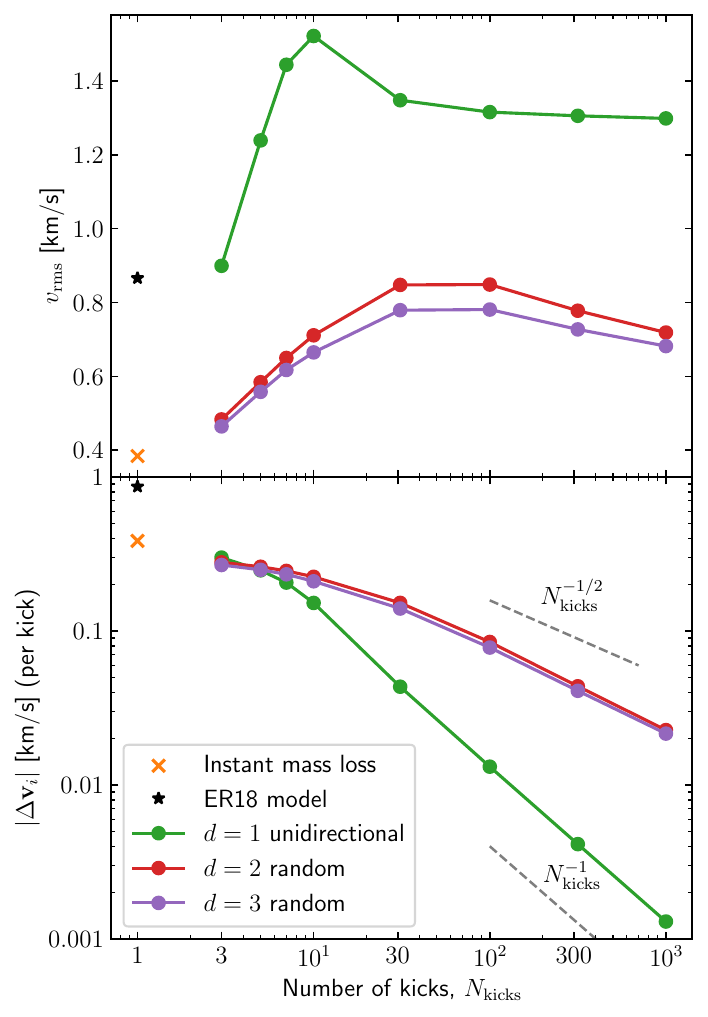}
    \caption{\textbf{Top}: Best-fitting cumulative root-mean-square kick velocities, $v_\mathrm{rms}$, as a function of the number of kicks, $N_\mathrm{kicks}$, for a single kick, unidirectional kick, and random kicks in two and three dimensions. We fit $v_\mathrm{rms}$ to the separation data of \citet{ElBadry2018}, abbreviated as ER18, for each kick model and $N_\mathrm{kicks}$. We also show the $v_\mathrm{rms}$ result from the ER18 model, where there is a single kick accompanied by adiabatic mass loss. \textbf{Bottom}: The corresponding root-mean-square velocity change per kick, $v_\mathrm{rms}/N_\mathrm{kicks}$ for coherent recoil in $d=1$ and $v_\mathrm{rms}/\sqrt{N_\mathrm{kicks}}$ for the random-kick models in $d=2, 3$. At large $N_\mathrm{kicks}$, these relations produce the scalings $N_\mathrm{kicks}^{-1}$ and $N_\mathrm{kicks}^{-1/2}$, respectively.}
    \label{fig:vkick}
\end{figure}

We now discuss the outcomes of the fitting process, applied to each of the natal-kick models considered in this work (single-kick, and $d=1,\,2,\,3$). In Fig.~\ref{fig:vkick}, we show the best-fit $\vrms$ (top) as well as the corresponding $\dv$ (bottom) as a function of $\nks$ for all models. 

For the single kick ($\nks=1$) with instant mass loss we obtain a best-fit $\vrms\approx0.4$ km/s.
For the single kick with \textit{adiabatic} mass loss, \citet{ElBadry2018} found that $v_\mathrm{rms} \approx 0.9$ km/s explains the separation data reasonably well, which we include as-reported without the fitting procedure used for all other models.
As $\nks$ increases, the best-fit $\vrms$ changes rapidly. For each of the models, a transient phase occurs where the best-fit $\vrms$ grows between $\nks=3$ and $30$, the most extreme example being the $d=1$ unidirectional kicks case. Following this initial adjustment, all models stabilize and approach what appear to be asymptotic values of $\vrms\approx 0.7$ km/s for both of the random walk models and $\approx1.3$ km/s for the $d=1$ recoil.
Additional simulations with higher $\nks$ can confirm these asymptotic values, but they become computationally prohibitive for us. Nonetheless, we later find that the $\nks$ in the range shown exhibit good agreement with observations (Fig. \ref{fig:ecc_chi2}), so we expect this range to be sufficient for our purposes.

In the bottom panel, two scalings appear for the many-kick scenarios, one for the $d{=}1$ unidirectional kicks, and another for the random walks. Because the high $\nks$ limit tends toward a (nearly) constant value for the best-fit $\vrms$ in all models, we can easily explain the cause of these scalings. Keeping both $\vrms$ and $v_{{\rm total},1d}$ constant, \eqref{eq:dv1d} indicates that the $d=1$ $\dv$ should scale as $\nks^{-1}$, and \eqref{eq:dvrw} indicates that the $d=2,3$ $\dv$ should scale as $\nks^{-1/2}$. Indeed, these are the scalings obtained for $\dv$ from the simulation results in Fig. \ref{fig:vkick}.

\begin{figure}
    \centering
    \includegraphics[width=1.\linewidth]{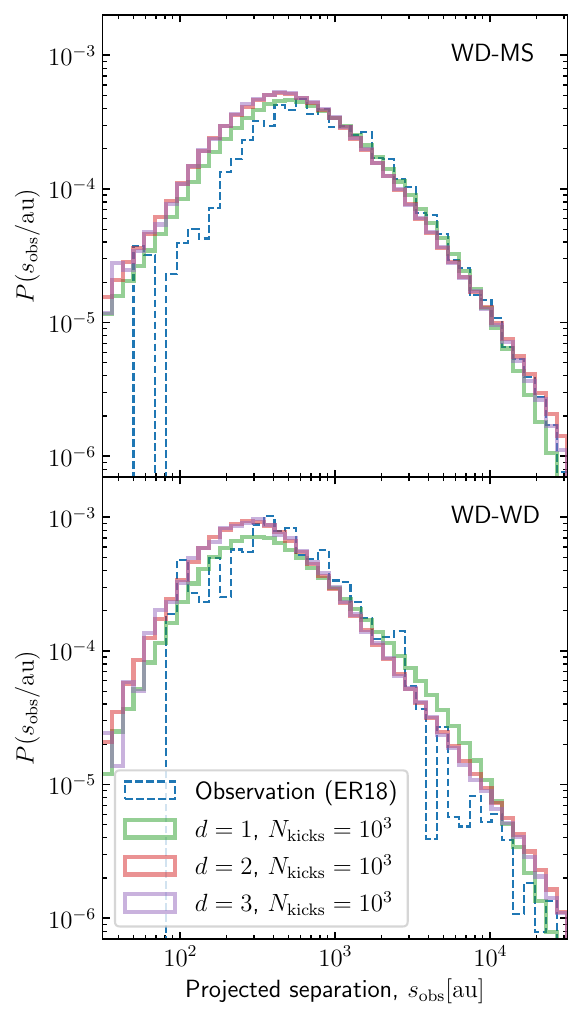}
    \caption{Both panels show the observed separation ($s_\mathrm{obs}$) distribution for WD--MS binaries (\textbf{top}) and WD--WD binaries (\textbf{bottom}). The dashed line shows the observational data from \citet{ElBadry2018}, denoted as ER18. The various colors show the mock observed data from binary population synthesis models with different kick types and numbers of kicks. We have applied the completeness correction (Appendix \ref{sec:mock}) to directly compare simulations to observations. We find that any kick type and number of kicks can fit the observed distribution well. }
    \label{fig:sobs}
\end{figure}

Fig.~\ref{fig:sobs} compares the observed separation distributions with mock observations for a few representative kick models.
We apply biases (cf. Appendix \ref{sec:mock}) to the output of the binary population synthesis models in order to mimic those present in the observational data. In the top panel, we show the comparison of the observed and synthesized $s_{\rm obs}$ distributions for WD--MS binaries. Given the limited precision of the observed distribution, visual comparison cannot reliably be used to rule out any of these models. The same is true of the WD--WD binary distributions (bottom panel), perhaps to an even greater extent, as the limited number of observed wide binaries contribute significantly to noise levels in the high-$s_{\rm obs}$ tail of the distribution.

\begin{figure}
    \centering
    \includegraphics[width=1.\linewidth]{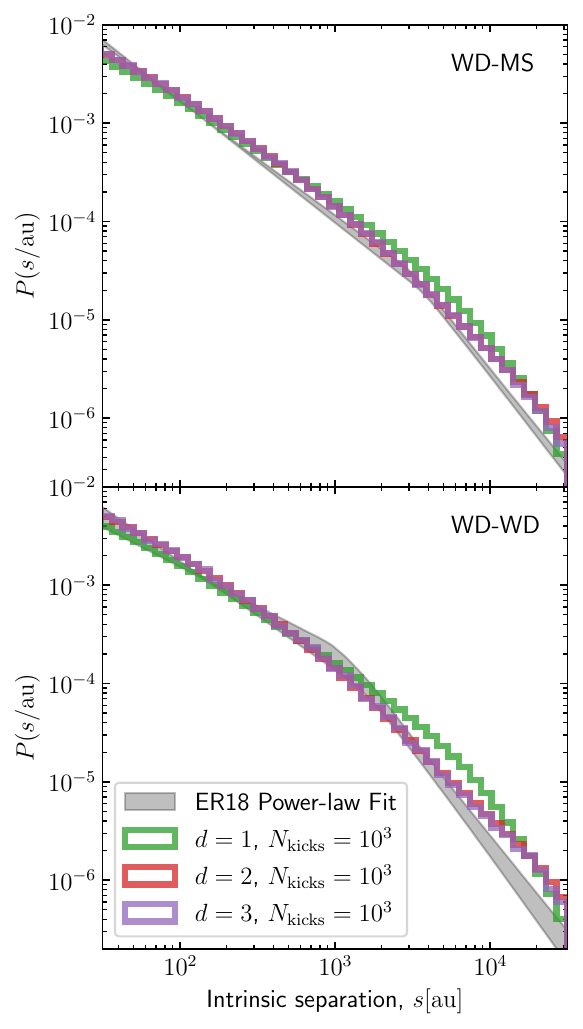}
    \caption{Both panels show the binary intrinsic separation, $s$, distribution for WD--MS binaries (\textbf{top}) and WD--WD binaries (\textbf{bottom}). The shaded region shows the broken power-law fit from \citet[][denoted as ER18 in the legend]{ElBadry2018}, which is a model itself. The other curves show the intrinsic separation of binaries from our binary population synthesis model, with various kick types and number of kicks. We find that our binary population synthesis model for $d=2,3$ recovers the power-law break.}
    \label{fig:strue}
\end{figure}

Unfortunately, a direct, de-biased comparison of the observed and synthesized intrinsic separation distributions is not as easily obtained as the biased, $s_{\rm obs}$ distributions are. Nonetheless, we may still compare our synthesized intrinsic separation ($s$) distributions to the broken power-law fit obtained in \citet{ElBadry2018} for the $s$ distribution of the observed binaries. We emphasize before doing so, however, that the broken power-law is strictly a fit to the observed data obtained via a maximum-likelihood procedure that is virtually identical to the one that we use. It is not the actual $s$ distribution of the observed binaries. With that said, we see that while all the models remain similar, subtle qualitative differences begin to arise. The random-walk models take on an appearance more reminiscent of a broken-power-law, while the $d=1$ unidirectional kicks model leads to a more rounded distribution. The high-$s$ tail of the $d=3$ random-walk model features a power-law tail that we measure to have an index of $\approx-1.7$. We note that this is somewhat shallower than the power-law tail obtained by the \citet{ElBadry2018} fit, $\approx-2$. We discuss the likely sources of this discrepancy in \S\ref{sec:comp}.

\section{The Eccentricity Distribution}\label{sec:ecc}

For each value of $\nks$ and $d$, the separation fitting process yielded a preferred $\vrms$. As seen from Fig. \ref{fig:sobs}, however, it is still unclear which model is best, and for what value of $\nks$. Here, we describe an additional comparison process where we use observed binary eccentricities to further narrow the scope of potential mass-loss/natal-kick models.

\subsection{Comparing with Observation}

The eccentricity distribution of a collection of observed binaries can be constrained via the $v-r$ angle technique which uses the angle between the binary's separation vector and relative velocity vector projected onto the plane of the sky \citep{Tokovinin1998, Tokovinin2016}.
Note that individual eccentricity measurement of a particular system cannot be done with this technique without knowing the inclination \citep{Hwang2022}.
Therefore it can only be used to constrain a \textit{distribution} of eccentricities.
The technique was previously used to measure the power-law index of the eccentricity distribution of MS--MS binaries as a function of their separation~\citep{Hwang2022}.
Applying this technique to the WD--MS and WD--WD binaries in the \textit{Gaia} DR3 catalog of \citet{ElBadry2021}, \citet{hwang2025} fit $P(e)\propto e^\alpha$ to the observed binary distributions and report the following three measurements of $\alpha$ in bins of projected separation:
\begin{enumerate}
    \item WD--MS with separations $[10^2, 10^3]$ au have $\alpha_1 = 0.8^{+0.1}_{-0.1}$,
    \item WD--MS with separations $[10^3, 3~000]$ au have $\alpha_2 = 1.0^{+0.2}_{-0.2}$,
    \item WD--WD with separations $[10^2, 10^3]$ au have $\alpha_3 = 0.2^{+0.3}_{-0.2}$.
\end{enumerate}

Using the previously found best-fit kick strengths, we can calculate $\alpha$ as a function of semi-major axis from our synthesized populations for each $\nks$ and $d$, and compare the results against these observations by calculating the $\chi^2$. When performing the power-law fit for $\alpha$, we only fit the eccentricity distribution up to $e_\mathrm{max}\in\{0.7,0.8,0.9,1.0\}$ and report the resultant median $\alpha$ and 16th--84th percentile range of the four fitted values of $\alpha$.
Having $e_\mathrm{max}$ is necessary because our potential CE removal criterion distorts the high-eccentricity tail, making the fitted index sensitive to the upper cutoff, which was also experienced by \citet{hwang2025}.
Additional details on the statistical procedure, including the importance and effects of $e_\mathrm{max}$, for this section can be found in Appendix \ref{appendix:ecc}.
Finally, we emphasize that in what follows, the uncertainties presented for the data are simply the $e_\mathrm{max}$ 16th--84th percentile ranges, not true statistical uncertainties or numerical errors.

\subsection{Results}

\begin{figure}
    \centering
    \includegraphics[width=1.\linewidth]{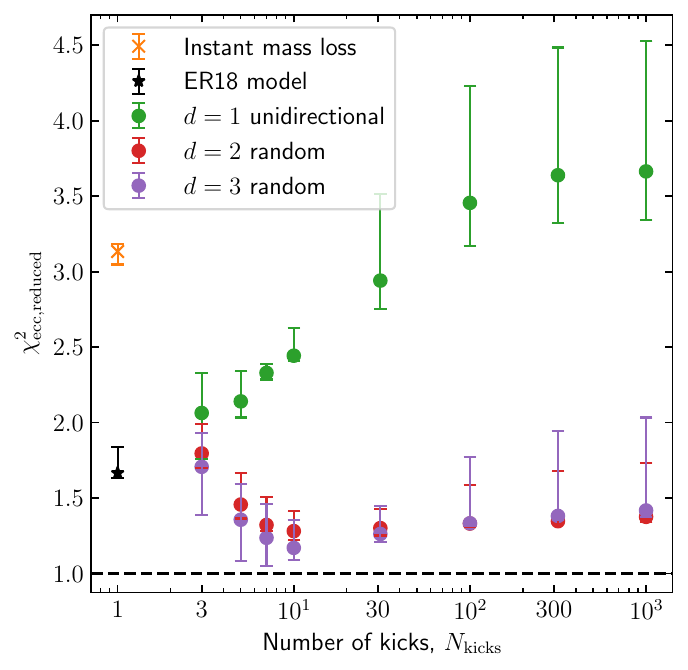}
    \caption{Reduced $\chi^2$ comparing the three simulated eccentricity indices with the observational constraints of \citet{hwang2025}, as a function of the number of kicks and the kick model. Colors distinguish the kick models. The single impulsive and unidirectional kick models give poorer overall agreement than the random-kick models. The single impulsive kick and unidirectional kick models cannot reproduce the observed eccentricity distributions of WD--MS and WD--WD wide binaries. We follow \citet{hwang2025} in how we are reporting the uncertainty bars in $\alpha$. Here, the uncertainty bars represent the sensitivity of $\alpha$ to the maximum eccentricity truncation, $e_\mathrm{max}$. These are not the statistical or simulation uncertainties. }
    \label{fig:ecc_chi2}
\end{figure}

\begin{figure}
    \centering
    \includegraphics[width=1.\linewidth]{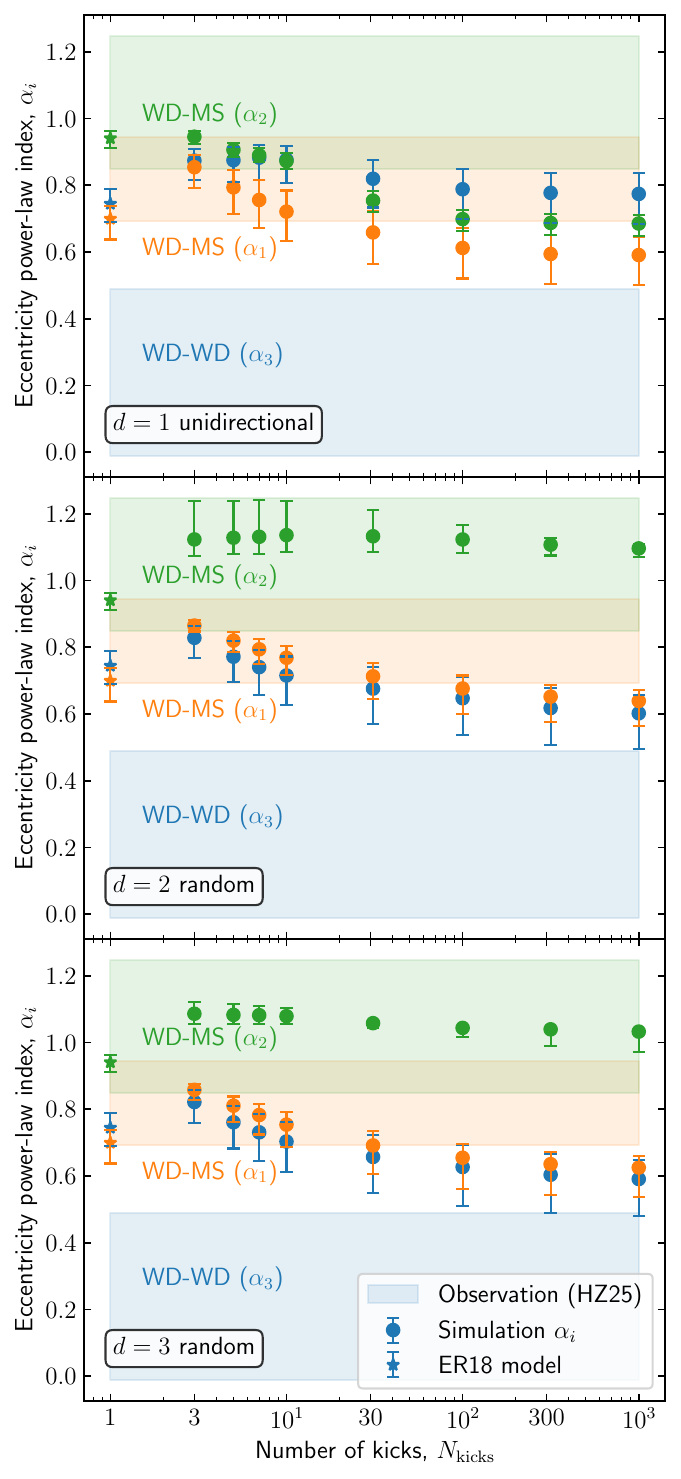}
    \caption{Best-fitting eccentricity power-law indices, $\alpha$ in $P(e)\propto e^\alpha$, as a function of the number of kicks. Each panel corresponds to a kick model. Colors differentiate the three combinations of binary type and projected-separation bin. Shaded regions show the observational constraints of \citet{hwang2025}, abbreviated as HZ25 in the legend. Points show the median simulated index across four upper eccentricity cutoffs, and bars span the corresponding 16th to 84th percentiles, quantifying sensitivity to $e_\mathrm{max}$.}
    \label{fig:alpha}
\end{figure}

\begin{figure}
    \centering
    \includegraphics[width=1.\linewidth]{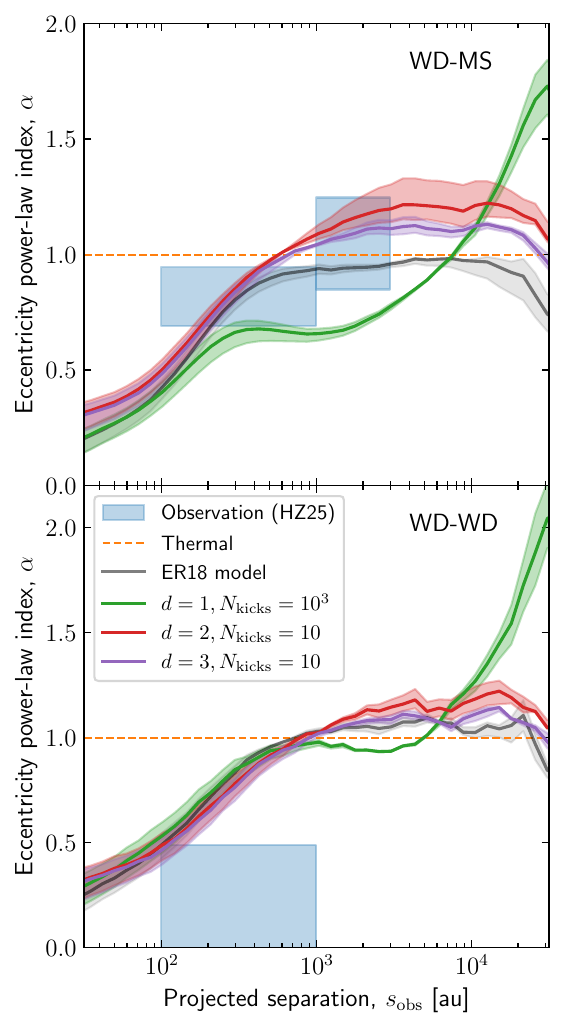}
    \caption{Eccentricity power-law index, $\alpha$ in $P(e)\propto e^\alpha$, as a function of projected separation, $s_\mathrm{obs}$, for WD--MS (\textbf{top}) and WD--WD binaries (\textbf{bottom}). Blue shaded regions show the observational constraints of \citet{hwang2025}, abbreviated as HZ25. Colored curves show the model predictions, with shading indicating sensitivity to the upper eccentricity cutoff, $e_\mathrm{max}$. The unidirectional kick is chosen with high $\nks = 10^3$ to contrast with the low $\nks$ limit from the ER18 model. The two- and three-dimensional random-kick curves use $N_\mathrm{kicks}=10$ and reproduce the WD--MS constraints most closely. The $\alpha$ results for the ER18 model are also shown. None of the considered models reproduces the WD--WD constraint.}
    \label{fig:ecc_sep}
\end{figure}

Fig.~\ref{fig:ecc_chi2} shows the reduced $\chi^2$ obtained for each best-fit $\vrms$ when comparing the indices predicted by the different kick models with the three observational constraints.
We find that the instantaneous mass-loss model with one impulsive kick performs very poorly.
For brevity, then, we do not discuss this model in subsequent analyses.
Next, the single-kick adiabatic mass-loss model by \citet{ElBadry2018}, hereafter ER18 model, achieves much better eccentricity agreement with data.
The two- and three-dimensional random-kick models with $N_\mathrm{kicks}\gtrsim10$ agree best with data.
For these models, $\chi^2$ is generally lower for $\nks \geq 10$ than it is for the few-kick cases, but we cannot report any definitive preferred $\nks$ value given $\chi^2$ and its uncertainties.

The $\chi^2$ results above can be understood by comparing them with Fig. \ref{fig:alpha}, which shows the best-fitting $\alpha$ values as a function of the number of kicks with separate panels for each of our kick models.
The observational constraints from \citet{hwang2025} are shown as shaded, colored areas.
The three different colors correspond to $\alpha_1$ (orange), $\alpha_2$ (green), and $\alpha_3$ (blue).
The models that explain the observational constraints are $d=2,3$ at $N_\mathrm{kicks}\approx10$.
As the number of kicks increases, $\alpha_2$ remains consistent with the observations, $\alpha_1$ moves outside the observed range, and $\alpha_3$ lowers but still cannot satisfy the WD--WD observational constraint.
The $d=1$ unidirectional kicks perform worst, since the ordering of their three $\alpha$ values does not match observations for any $\nks$. 
To a lesser extent, the ER18 model suffers from the same effect with $\alpha_1$ (orange) and $\alpha_3$ (blue) flipped in ordering.
We find that no model produces the $\alpha_3$ observed value.
The origin of the discrepancy in the WD--WD index $\alpha_3$ remains unclear; the limited WD--WD eccentricity sample and the comparatively large uncertainty reported by \citet{hwang2025} limit its interpretation.
We discuss some possible reasons for the apparent disagreement in \S\ref{sec:caveats}.

In Fig. \ref{fig:ecc_sep}, we show the eccentricity power-law index $\alpha$ as a function of observed separation for both WD--MS and WD--WD binaries from our simulations for representative choices of $\nks$.
For the unidirectional $d=1$ model, we show $\nks=10^3$ to compare the kick results in the high-$\nks$ (``smooth'') limits to contrast with the ER18 single kick model's results.
For the random-walk models, we show the distributions for $\nks=10$ as representative examples.
This figure shows that our slow, unidirectional kick model gives poor agreement with the observed eccentricity indices for all cases.\footnote{Again, we note that the unidirectonal kick model implemented here is different than that from \citet{hwang2025, oconnor2026} (cf. \S\ref{sec:kickmodels}, \S\ref{sec:comparisons}), and the results here do not necessarily rule out the continuous recoil model in these papers.}
Notably, it stands in contrast to all other models beyond $10^{3.5}$ au, where it produces super-thermal distributions. 
Similar to Fig. \ref{fig:alpha}, we find that the ER18 model and the $d=2,3$ models with $\nks \approx 10$ produce eccentricity power-laws in agreement with observational constraints for WD--MS binaries.

The results in Fig. \ref{fig:ecc_sep} are also our predictions for future observations of wide-binary eccentricity distributions as a function of projected separations.
New eccentricity distribution constraints obtained from the $v-r$ technique on data from the upcoming \textit{Gaia} DR4 release could be compared with this figure.
These data, together with our predicted $\alpha$, might clarify further the current tension between models and observation of $\alpha_3$.

\section{Results for Common Envelope and Dissolved Binaries}\label{sec:cedis}

Each model for WD natal kicks that we have presented yields slightly different distributions for the binary orbital parameters. It is therefore expected that each model also yields different predictions for the fraction of binaries that are dissolved by WD natal kicks, as well as the fraction of binaries that are driven into common-envelope evolution. In this section, we present the results for these fractions using the best-fit values of $\vrms$ for each $\nks$ and $d$ (as well as the ER18 model).

\begin{figure}
    \centering
    \includegraphics[width=1.\linewidth]{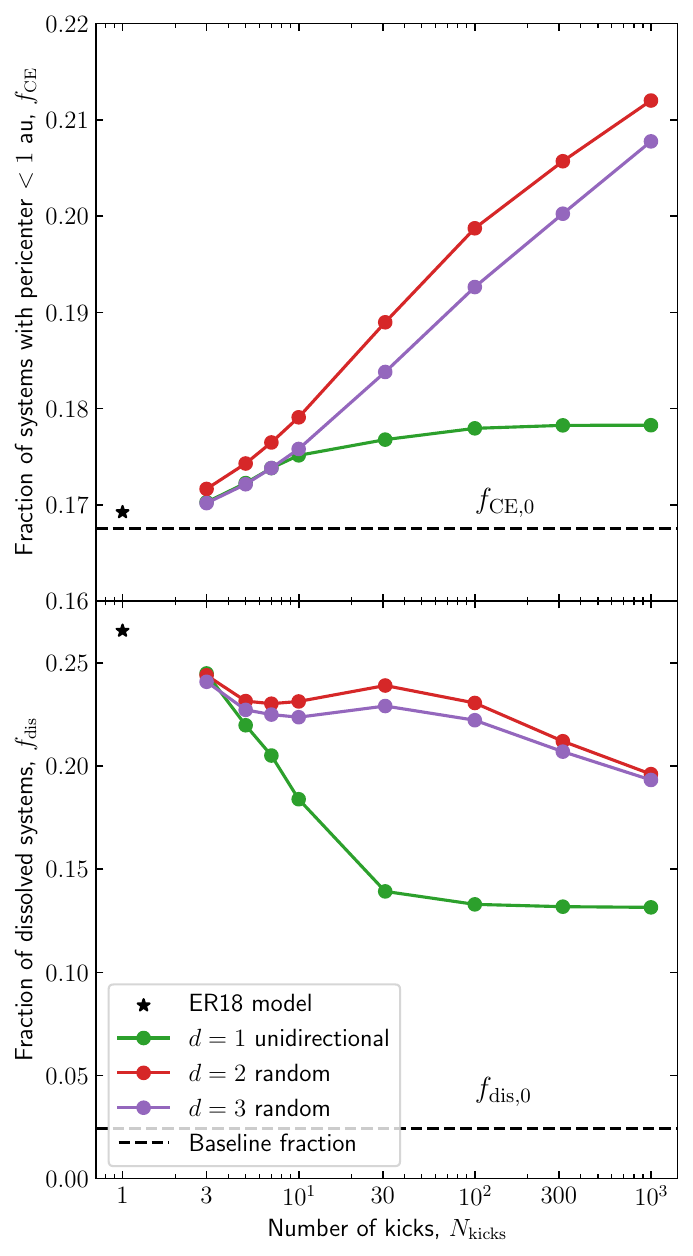}
    \caption{\textbf{Top}: Fraction of simulated binaries removed as potential common envelope systems after reaching pericenter distances below $1$ au. \textbf{Bottom}: Fraction of binaries classified as dissolved at the end of stellar evolution: unbound systems or those with semi-major axes above $10^5$ au, motivated by the stellar-flyby disruption estimate in \citet{hamilton2024b}.
    Fractions in both panels are relative to the initial simulated population capable of forming at least one WD within the age of the Galaxy. The dashed line in both panels marks the initial fraction of binaries satisfying these two respective cases. Colors distinguish the kick models.}
    \label{fig:fCEdis}
\end{figure}

The top panel of Fig. \ref{fig:fCEdis} shows the fraction of simulated binaries that reach pericenter distances below $1$ au and are removed as potential CE systems.
This fraction is calculated relative to the simulated population, which includes only binaries in which at least one star can undergo post-MS evolution within the age of the Galaxy.
The horizontal dashed line shows the fraction of systems whose initial pericenter distances are already below the adopted threshold, so that they could interact when a component becomes a red giant, before the TP-AGB mass-loss and kick events.
Approximately $17-21$\% of the simulated systems satisfy the CE criterion when $\nks \leq 10^{3}$, with kicks enhancing the baseline fraction of 17\% by only a few percent. The fractions for $d=2$ and $d=3$ yield the greatest changes in CE fraction, and are nearly equivalent to each other.
The $d = 1$ unidirectional model yields almost no change in the CE fraction.

The bottom panel of Fig.~\ref{fig:fCEdis} shows the fraction of simulated systems classified as dissolved if, by the end of stellar evolution, they are unbound ($e>1$) or have a semi-major axis greater than $10^5$ au.
The $10^5$ au threshold is approximately the distance where wide binaries are susceptible to dissolution by stellar flybys \citep{hamilton2024b}.
While this separation technically depends on the binary mass, the mass of the passing star, and the flyby distance, it does not increase by more than an order of magnitude.
In addition to flybys, \citet{Heisler1986} found that $10^5$ au is also roughly the Hill limit, beyond which the Galactic tide can dissolve binaries.
We find that approximately $13\%$ of binaries are classified as dissolved or susceptible to dissolution in the $d=1$ unidirectional kick model, consistent with \citet{hwang2025}.
Having randomly directed velocity kicks enhances the dissolution fraction to $\sim 20-25\%$, depending on the number of kicks incurred.

Of course, the fractions reported here depend on the adopted removal criteria and should be interpreted accordingly.
The CE criterion identifies potentially interacting systems, but does not determine whether those systems subsequently survive, merge, or leave the wide-binary population.
For the dissolution fraction, we use the conservative lower value estimate of $10^5$ au, which can be scaled using Eq. 14 in \citet{hamilton2024b} for stronger flybys or higher mass binaries.
The $10^5\mathrm{~au}$ threshold can also be scaled with other binary masses using Eq. 8 in \citet{Heisler1986} for dissolution via Galactic tide.
Therefore, we emphasize that these fractions characterize the removal criteria adopted in our simulations and do not \textit{directly} predict the fraction of binaries permanently destroyed during post-MS evolution.

\section{Discussion and Conclusion}\label{sec:disc}

\subsection{Comparison with Paper I}\label{sec:comp}

The approach that we have taken here to investigate the random-walk paradigm of WD natal kicks is very distinct from that of \pI. In \pI, we adopted a Fokker-Planck framework to describe the evolution of a smooth probability distribution of binary orbital parameters subject to diffusion and drift from randomly directed kicks and mass loss. In that framework, we assume $\nks \gg1$ so that we may reasonably take each kick-induced change in the orbital energy and eccentricity to be small relative to their magnitudes. Our eccentricity comparison favors random-kick models with $N_\mathrm{kicks}\gtrsim10$
(Fig.~\ref{fig:ecc_chi2}).
We therefore find that the results here qualitatively support the multiple-kick framework for post-MS evolution, and the use of the Fokker-Planck approach as a way to understand its dynamical consequences.

The primary aim of \pI~is distinct from this work. In \pI, we focused on the nuanced dynamical physics of the random-walk process, rather than attempting to obtain detailed information about the strength of the kicks or their quantity, as we do here. As a result, many of our primary conclusions focused on qualitative properties of the resultant orbital-parameter distributions. 
One of the most significant aspects of that work was a first-principles demonstration that random-walk WD natal kicks necessarily induce a $-2$ power-law tail in the semi-major axis distribution of wide binaries (which, as noted in \citealt{ElBadry2018}, corresponds to a similar power law tail in the intrinsic separation distribution).
This seems to be strong evidence in favor of the random-walk paradigm, because \citet{ElBadry2018} reported a $-2$ power-law tail as the best fit to their intrinsic binary-separation distributions. Here, we also find a power-law tail in the separation distributions of WD binaries, which although slightly shallower than the data (power-law index of ${\approx}-1.7$), largely confirms the result that we obtained through analytic theory in \pI. The most likely cause for this slight difference is that in \pI, we only considered populations of a single mass and mass ratio in each distribution. Here, we have a broad combination of masses and mass ratios, all of which are modified by varying amounts as a result of the kicks. We suspect, then, that the slight flattening with respect to the expected $-2$ power-law is a consequence of the variance in the binary parameters, and that a similar convolution of all Fokker-Planck distributions at many masses and mass ratios might also muddle the clarity of the $-2$ power-law tail. We note that this also hints at the importance of accurate initial period distributions as a function of mass; a different prescription for the relationship between binary mass and initial orbital periods could yield a slightly different final separation distribution. Nonetheless, the measurement of a similar power-law tail to that expected through analytic theory and observations is an encouraging sign for the applicability of a random-walk approach to WD natal kicks.

Although our focus in \pI~was more qualitative than this work, there were still numerical results that can be compared to the values obtained here. In particular, we predicted that for $\vrms \sim 1$ km/s, the eccentricity distribution should undergo a rapid transition to a thermal distribution near $a\sim 10^3$ au. Here, we can clearly see that this occurs in the top panel of Fig. \ref{fig:ecc_sep}. There is some overshoot of the thermal distribution, however, we note that this is likely due to the lower number of kicks shown ($\nks=10$), which makes the dynamical modifications less gentle than in a Fokker-Planck approach. In Fig. \ref{fig:alpha}, it can be seen that for $\nks =10^3$ the $d=3$ eccentricity distribution becomes closer to thermal beyond $s\sim 10^3$ au.
In \pI, we also calculated the fraction of binaries that would eventually be dissolved by stellar flybys and the Galactic tide, for varying mass, mass ratio, and kick strengths. We found that, for the most massive binaries subject to kicks of $\vrms=0.5$ km/s, only 5\% of binaries are dissolved, while for the lowest mass binaries with $\vrms=1$ km/s, 21\% of binaries are dissolved. In this work, we find that roughly 20\% of all binaries are dissolved for the highest-$\nks$ scenario. Given that the stellar initial-mass function is heavily weighted toward lower mass stars, the result of $\approx 20\%$ here indicates satisfactory agreement between the Fokker-Planck treatment and the binary population synthesis approach.

\subsection{Comparison with Previous Work}\label{sec:comparisons}

In Figs. \ref{fig:ecc_chi2}, \ref{fig:alpha}, \ref{fig:ecc_sep}, we show that our unidirectional kick model is disfavored as it cannot constrain both the observed separation and eccentricity distributions from \citet{ElBadry2018, hwang2025}.
We cannot necessarily rule out the recoil model used in \citet{hwang2025, oconnor2026} as there are differences between this paper and theirs in the implementation of the kick and mass loss model.
As previously discussed in \S\ref{sec:kickmodels}, our kicks are independent of mass loss and occur discretely, in contrast with previous literature.

That said, we emphasize that our approach here provides stronger constraints on the kick strength compared to other works. This is owed in particular to the fact that we directly fit the optimal kick strength against the separation data and then compare the eccentricity distributions. 
In \citet{ElBadry2018}, the kick strength is qualitatively constrained by the observed separation data (see Fig. 11 in their work).
In \citet{hwang2025}, the separation data is used to qualitatively constrain the recoil kick strength, although the observational eccentricity data is reproduced quantitatively (see Fig. 7 and Sec. 4.2 in their paper).
In no previous work is there a direct quantitative evaluation of the agreement between models and observational data for \textit{both} the separation and eccentricity distributions.
Furthermore, to our knowledge, this work offers the first side-by-side comparison between distinct kick models, which enables us to probe more precisely the basic nature of white dwarf kicks.

\subsection{Caveats}\label{sec:caveats}

There are several caveats with the methodology implemented in this paper.

First, we fit kick strengths to the separation data and compare the resulting eccentricity predictions with the constraints of \citet{hwang2025}.
A joint analysis would need to combine thousands of individual projected separations with only three population-level eccentricity indices, $\alpha$.
A hierarchical Bayesian framework could accommodate both types of constraints to infer the population-level kick parameters \citep[e.g.,][]{Poon2025, Poon2026}; we defer such a task to future work.

Second, our mass-loss prescription assumes that the star loses the same amount of mass in each event.
A more detailed prescription could allow the mass-loss rate to vary with time.
Although such a prescription could be incorporated into our model, the present treatment and \citep{hwang2025} already reproduces the main observed trends.

Third, we neglect the effects of external perturbations on binary orbital evolution, such as those from the Galactic tide, stellar flybys, and interactions with molecular clouds.
The Galactic tide can induce significant eccentricity changes and drive binaries to small pericenter distances, particularly at separations of order a few $\times 10^4$ au \citep{Heisler1986, modak2023, Pham2024}.
Stellar flybys can modify both the eccentricity and semi-major axis distributions.
The characteristic Galactic-tide timescale of approximately $10^8$ yr and the typical interval between strong stellar flybys are both much longer than the TP-AGB lifetime \citep[cf.][]{Heisler1986, Zink2020, Brown2022}, so we expect these perturbations to be negligible during this phase.
Over longer timescales, however, both processes can reshape the orbital distributions after WD formation, especially during the interval between the WD--MS and WD--WD stages.
The Galactic tide could also increase the fraction of binaries reaching the CE threshold in Fig.~\ref{fig:fCEdis} by driving binaries to high eccentricities.
For discussions of how these processes shape wide-binary populations, we refer the reader to \citet{Heisler1986, modak2023} for the Galactic tide, \citet{jiang2010, hamilton2024b} for stellar flybys, and \citet{weinberg1987} for molecular clouds.

Fourth, we do not include additional physics for close encounters, such as CE or tidal effects.
As discussed by \citet{oconnor2026}, tidal interactions in close binaries can modify the eccentricity distribution and circularize some binaries.
CE evolution can also substantially alter binary orbits \citep{Ivanova2013, Grondin2024, Grondin2026}.
Both of these effects can potentially explain the WD--WD $\alpha_3$ discrepancy, if the evolution of these binaries is more susceptible to these effects. 
Including these additional physical processes is computationally costly and beyond the scope of this work.

Despite these limitations, our binary population synthesis model reproduces the main observed trends in the separation and eccentricity distributions of WD--MS binaries.
Additional physical processes may, however, be needed to explain the WD--WD eccentricity power-law index $\alpha_3$ (Figs. \ref{fig:alpha}, \ref{fig:ecc_sep}, \ref{fig:eobs}).
The low observed value of $\alpha_3$ implies that these WD--WD binaries are more circularized than expected.
Binary evolution that includes the aforementioned missing physics may resolve this issue.

\subsection{Conclusion}\label{sec:conclusion}

We develop a binary population synthesis model incorporating post-main-sequence mass loss, single impulsive kicks, unidirectional kicks, and random kicks in two and three dimensions.
All tested kick models reproduce the separation data after fitting their kick strengths, with a best-fitting cumulative root-mean-square velocity of approximately $0.7$ km/s for the multiple-random-kick cases (Fig.~\ref{fig:vkick}).
The eccentricity observational constraints of \citet{hwang2025} favor random-kick models having at least 10 kicks, with WD--MS binaries best reproduced.
Among simulated binaries capable of undergoing post-MS evolution, $17$--$20\%$ enter common-envelope evolution, and approximately $20$--$25\%$ are dissolved.

The population synthesis results are broadly consistent with the Fokker-Planck analytic predictions of \pI.
In addition, we identify the limitations of our model and discuss possible improvements.
Despite these limitations, the model reproduces the main observed trends.
However, the WD--WD eccentricity distribution remains a challenge.
The upcoming \textit{Gaia} DR4 release could refine these constraints and strengthen the connection between wide WD binary dynamics and post-main-sequence evolution.

During the preparation of this manuscript, we became aware of the concurrent work of \citet{Fuller2026}. In their paper, a substantially different approach is taken to the problem of random-walk white dwarf natal kicks.
A comparison of the resultant dynamical physics can be found in \pI.
As for the inferred kick properties, our maximum-likelihood approach obtains a root-mean-square kick strength (for $d=3$, the closest analog to their treatment) of ${\approx}0.7$ km/s, which is quite close to their value of ${\approx}0.5$ km/s. 
Notably, when comparing with the observed eccentricity distributions, we find that models with at least $\nks\gtrsim10$ are most favored.
This number is a lower bound, and could be interpreted as being consistent with the conclusion of \citet{Fuller2026} of $\nks\sim 10^4$.
We find this tentative agreement encouraging: both approaches suggest that multiple random-walk kicks occur during AGB evolution.

\section*{Acknowledgments}
We thank the anonymous referee for comments that improved this manuscript.
This work benefited greatly from conversations with Shaunak Modak, Hanno Rein, Robert Ewart, Thomas Foster, Matthew Kunz, Ann-Marie Madigan, and Drake Miller.
We thank Nadia Zakamska, Chris O'Connor, and Yanqin Wu for inspiring us to study white dwarf kicks.

This work utilized the Alpine High-Performance Computing resource at the University of Colorado Boulder. Alpine is jointly funded by the University of Colorado Boulder, the University of Colorado Anschutz, Colorado State University, and the National Science Foundation (award 2201538).
This work used ACES at Texas A\&M High Performance Research Computing through allocation \#PHY260085 from the Advanced Cyberinfrastructure Coordination Ecosystem: Services \& Support (ACCESS) program, which is supported by U.S. National Science Foundation grants \#2138259, \#2138286, \#2138307, \#2137603, and \#2138296.

Support for S.M. was provided by NASA Astrophysics Program grant 80NSSC22K0828.
D.P. acknowledges support from the McCray Postdoctoral Fellowship at the University of Colorado Boulder.

\section*{Data Availability}
The data underlying this article will be shared on reasonable request to the corresponding authors.

\bibliography{references}{}

@article{hundhausen1993,
author = {Hundhausen, A. J.},
title = {Sizes and locations of coronal mass ejections: SMM observations from 1980 and 1984-1989},
journal = {Journal of Geophysical Research: Space Physics},
volume = {98},
number = {A8},
pages = {13177-13200},
doi = {https://doi.org/10.1029/93JA00157},
url = {https://agupubs.onlinelibrary.wiley.com/doi/abs/10.1029/93JA00157},
eprint = {https://agupubs.onlinelibrary.wiley.com/doi/pdf/10.1029/93JA00157},
year = {1993}
}

@ARTICLE{paper1,
       author = {{Majeski}, Stephen and {Pham}, Dang and {Ryu}, Taeho and {Tomassini}, Gabriel and {Chiavassa}, Andrea},
        title = "{White Dwarf Natal Kicks as Velocity-Space Random Walks. I. Fokker-Planck Theory}",
      journal = {arXiv e-prints},
         year = 2026,
        month = sep,
          eid = {arXiv:2609.20940},
        pages = {arXiv:2609.20940},
          doi = {10.48550/arXiv.2609.20940},
archivePrefix = {arXiv},
       eprint = {2609.20940},
 primaryClass = {astro-ph.SR},
       adsurl = {https://ui.adsabs.harvard.edu/abs/2026arXiv260920940M}
}

@ARTICLE{hamilton2024b,
       author = {{Hamilton}, Chris and {Modak}, Shaunak},
        title = "{Eccentricity dynamics of wide binaries - II. The effect of stellar encounters and constraints on formation channels}",
      journal = {\mnras},
         year = 2024,
        month = aug,
       volume = {532},
       number = {2},
        pages = {2425-2440},
          doi = {10.1093/mnras/stae1654},
archivePrefix = {arXiv},
       eprint = {2311.04352},
 primaryClass = {astro-ph.GA},
       adsurl = {https://ui.adsabs.harvard.edu/abs/2024MNRAS.532.2425H}
}

@ARTICLE{dotter2016,
       author = {{Dotter}, Aaron},
        title = "{MESA Isochrones and Stellar Tracks (MIST) 0: Methods for the Construction of Stellar Isochrones}",
      journal = {\apjs},
         year = 2016,
        month = jan,
       volume = {222},
       number = {1},
          eid = {8},
        pages = {8},
          doi = {10.3847/0067-0049/222/1/8},
archivePrefix = {arXiv},
       eprint = {1601.05144},
 primaryClass = {astro-ph.SR},
       adsurl = {https://ui.adsabs.harvard.edu/abs/2016ApJS..222....8D}
}

@ARTICLE{duquennoy1991,
       author = {{Duquennoy}, A. and {Mayor}, M.},
        title = "{Multiplicity among Solar Type Stars in the Solar Neighbourhood - Part Two - Distribution of the Orbital Elements in an Unbiased Sample}",
      journal = {\aap},
         year = 1991,
        month = aug,
       volume = {248},
        pages = {485},
       adsurl = {https://ui.adsabs.harvard.edu/abs/1991A&A...248..485D}
}

@ARTICLE{fellhauer2003,
       author = {{Fellhauer}, M. and {Lin}, D.~N.~C. and {Bolte}, M. and {Aarseth}, S.~J. and {Williams}, K.~A.},
        title = "{The White Dwarf Deficit in Open Clusters: Dynamical Processes}",
      journal = {\apjl},
         year = 2003,
        month = sep,
       volume = {595},
       number = {1},
        pages = {L53-L56},
          doi = {10.1086/379005},
archivePrefix = {arXiv},
       eprint = {astro-ph/0308261},
 primaryClass = {astro-ph},
       adsurl = {https://ui.adsabs.harvard.edu/abs/2003ApJ...595L..53F}
}

@ARTICLE{hwang2025,
       author = {{Hwang}, Hsiang-Chih and {Zakamska}, Nadia L.},
        title = "{White Dwarfs in Wide Binaries: The Strong Effects of Stellar Evolution and Mass Loss}",
      journal = {\apj},
         year = 2025,
        month = oct,
       volume = {991},
       number = {2},
          eid = {226},
        pages = {226},
          doi = {10.3847/1538-4357/adfa1c},
archivePrefix = {arXiv},
       eprint = {2508.08364},
 primaryClass = {astro-ph.SR},
       adsurl = {https://ui.adsabs.harvard.edu/abs/2025ApJ...991..226H}
}

@ARTICLE{oconnor2026,
       author = {{O'Connor}, Christopher E.},
        title = "{The Fate of Gaia's Wide Binaries: Interplay of White-dwarf Recoil and Tidal Interactions}",
      journal = {\apj},
         year = 2026,
        month = feb,
       volume = {998},
       number = {2},
          eid = {280},
        pages = {280},
          doi = {10.3847/1538-4357/ae3b26},
archivePrefix = {arXiv},
       eprint = {2509.08880},
 primaryClass = {astro-ph.SR},
       adsurl = {https://ui.adsabs.harvard.edu/abs/2026ApJ...998..280O}
}

@ARTICLE{jiang2010,
       author = {{Jiang}, Yan-Fei and {Tremaine}, Scott},
        title = "{The evolution of wide binary stars}",
      journal = {\mnras},
         year = 2010,
        month = jan,
       volume = {401},
       number = {2},
        pages = {977-994},
          doi = {10.1111/j.1365-2966.2009.15744.x},
archivePrefix = {arXiv},
       eprint = {0907.2952},
 primaryClass = {astro-ph.GA},
       adsurl = {https://ui.adsabs.harvard.edu/abs/2010MNRAS.401..977J}
}

@ARTICLE{weinberg1987,
       author = {{Weinberg}, Martin D. and {Shapiro}, Stuart L. and {Wasserman}, Ira},
        title = "{The Dynamical Fate of Wide Binaries in the Solar Neighborhood}",
      journal = {\apj},
         year = 1987,
        month = jan,
       volume = {312},
        pages = {367},
          doi = {10.1086/164883},
       adsurl = {https://ui.adsabs.harvard.edu/abs/1987ApJ...312..367W}
}

@ARTICLE{modak2023,
       author = {{Modak}, Shaunak and {Hamilton}, Chris},
        title = "{Eccentricity dynamics of wide binaries - I. The effect of Galactic tides}",
      journal = {\mnras},
         year = 2023,
        month = sep,
       volume = {524},
       number = {2},
        pages = {3102-3115},
          doi = {10.1093/mnras/stad2073},
archivePrefix = {arXiv},
       eprint = {2303.15531},
 primaryClass = {astro-ph.GA},
       adsurl = {https://ui.adsabs.harvard.edu/abs/2023MNRAS.524.3102M}
}

@ARTICLE{Ivanova2013,
       author = {{Ivanova}, N. and {Justham}, S. and {Chen}, X. and {De Marco}, O. and {Fryer}, C.~L. and {Gaburov}, E. and {Ge}, H. and {Glebbeek}, E. and {Han}, Z. and {Li}, X.-D. and {Lu}, G. and {Marsh}, T. and {Podsiadlowski}, P. and {Potter}, A. and {Soker}, N. and {Taam}, R. and {Tauris}, T.~M. and {van den Heuvel}, E.~P.~J. and {Webbink}, R.~F.},
        title = "{Common envelope evolution: where we stand and how we can move forward}",
      journal = {\aapr},
         year = 2013,
        month = feb,
       volume = {21},
          eid = {59},
        pages = {59},
          doi = {10.1007/s00159-013-0059-2},
archivePrefix = {arXiv},
       eprint = {1209.4302},
 primaryClass = {astro-ph.HE},
       adsurl = {https://ui.adsabs.harvard.edu/abs/2013A&ARv..21...59I}
}

@ARTICLE{maercker2012,
       author = {{Maercker}, M. and {Mohamed}, S. and {Vlemmings}, W.~H.~T. and {Ramstedt}, S. and {Groenewegen}, M.~A.~T. and {Humphreys}, E. and {Kerschbaum}, F. and {Lindqvist}, M. and {Olofsson}, H. and {Paladini}, C. and {Wittkowski}, M. and {de Gregorio-Monsalvo}, I. and {Nyman}, L.-A.},
        title = "{Unexpectedly large mass loss during the thermal pulse cycle of the red giant star R Sculptoris}",
      journal = {\nat},
         year = 2012,
        month = oct,
       volume = {490},
       number = {7419},
        pages = {232-234},
          doi = {10.1038/nature11511},
archivePrefix = {arXiv},
       eprint = {1210.3030},
 primaryClass = {astro-ph.SR},
       adsurl = {https://ui.adsabs.harvard.edu/abs/2012Natur.490..232M}
}

@ARTICLE{cui2026,
       author = {{Cui}, Yingzhen and {Wang}, Song and {Meng}, Xiangcun and {Liu}, Jifeng and {Ma}, Shuguo and {Zhao}, Weitao},
        title = "{Dynamical mass loss at the end of thermally pulsating asymptotic giant branch stars}",
      journal = {\aap},
         year = 2026,
        month = mar,
       volume = {707},
          eid = {A342},
        pages = {A342},
          doi = {10.1051/0004-6361/202556962},
archivePrefix = {arXiv},
       eprint = {2601.18194},
 primaryClass = {astro-ph.SR},
       adsurl = {https://ui.adsabs.harvard.edu/abs/2026A&A...707A.342C}
}

@ARTICLE{ElBadry2018,
       author = {{El-Badry}, Kareem and {Rix}, Hans-Walter},
        title = "{Imprints of white dwarf recoil in the separation distribution of Gaia wide binaries}",
      journal = {\mnras},
         year = 2018,
        month = nov,
       volume = {480},
       number = {4},
        pages = {4884-4902},
          doi = {10.1093/mnras/sty2186},
archivePrefix = {arXiv},
       eprint = {1807.06011},
 primaryClass = {astro-ph.SR},
       adsurl = {https://ui.adsabs.harvard.edu/abs/2018MNRAS.480.4884E}
}

@ARTICLE{Fuller2026,
       author = {{Fuller}, Jim},
        title = "{White Dwarf Kicks via Episodic Mass Ejection from Red Giant Stars}",
      journal = {arXiv e-prints},
         year = 2026,
        month = aug,
          eid = {arXiv:2608.07455},
        pages = {arXiv:2608.07455},
archivePrefix = {arXiv},
       eprint = {2608.07455},
 primaryClass = {astro-ph.SR},
       adsurl = {https://ui.adsabs.harvard.edu/abs/2026arXiv260807455F}
}

@ARTICLE{Heyl2007,
       author = {{Heyl}, Jeremy},
        title = "{Constraining white dwarf kicks in globular clusters}",
      journal = {\mnras},
         year = 2007,
        month = oct,
       volume = {381},
       number = {1},
        pages = {L70-L73},
          doi = {10.1111/j.1745-3933.2007.00369.x},
archivePrefix = {arXiv},
       eprint = {0706.0900},
 primaryClass = {astro-ph},
       adsurl = {https://ui.adsabs.harvard.edu/abs/2007MNRAS.381L..70H}
}

@ARTICLE{Davis2008,
       author = {{Davis}, D.~S. and {Richer}, H.~B. and {King}, I.~R. and {Anderson}, J. and {Coffey}, J. and {Fahlman}, G.~G. and {Hurley}, J. and {Kalirai}, J.~S.},
        title = "{On the radial distribution of white dwarfs in the globular cluster NGC 6397}",
      journal = {\mnras},
         year = 2008,
        month = jan,
       volume = {383},
       number = {1},
        pages = {L20-L24},
          doi = {10.1111/j.1745-3933.2007.00402.x},
archivePrefix = {arXiv},
       eprint = {0709.4286},
 primaryClass = {astro-ph},
       adsurl = {https://ui.adsabs.harvard.edu/abs/2008MNRAS.383L..20D}
}

@ARTICLE{Fregeau2009,
       author = {{Fregeau}, John M. and {Richer}, Harvey B. and {Rasio}, Frederic A. and {Hurley}, Jarrod R.},
        title = "{The Dynamical Effects of White Dwarf Birth Kicks in Globular Star Clusters}",
      journal = {\apjl},
         year = 2009,
        month = apr,
       volume = {695},
       number = {1},
        pages = {L20-L24},
          doi = {10.1088/0004-637X/695/1/L20},
archivePrefix = {arXiv},
       eprint = {0902.1166},
 primaryClass = {astro-ph.GA},
       adsurl = {https://ui.adsabs.harvard.edu/abs/2009ApJ...695L..20F}
}

@ARTICLE{Ohnaka2016,
       author = {{Ohnaka}, K. and {Weigelt}, G. and {Hofmann}, K.-H.},
        title = "{Clumpy dust clouds and extended atmosphere of the AGB star W Hydrae revealed with VLT/SPHERE-ZIMPOL and VLTI/AMBER}",
      journal = {\aap},
         year = 2016,
        month = may,
       volume = {589},
          eid = {A91},
        pages = {A91},
          doi = {10.1051/0004-6361/201628229},
archivePrefix = {arXiv},
       eprint = {1603.01197},
 primaryClass = {astro-ph.SR},
       adsurl = {https://ui.adsabs.harvard.edu/abs/2016A&A...589A..91O}
}

@ARTICLE{Ohnaka2017,
       author = {{Ohnaka}, K. and {Weigelt}, G. and {Hofmann}, K.-H.},
        title = "{Clumpy dust clouds and extended atmosphere of the AGB star W Hydrae revealed with VLT/SPHERE-ZIMPOL and VLTI/AMBER. II. Time variations between pre-maximum and minimum light}",
      journal = {\aap},
         year = 2017,
        month = jan,
       volume = {597},
          eid = {A20},
        pages = {A20},
          doi = {10.1051/0004-6361/201629761},
archivePrefix = {arXiv},
       eprint = {1611.04622},
 primaryClass = {astro-ph.SR},
       adsurl = {https://ui.adsabs.harvard.edu/abs/2017A&A...597A..20O}
}

@ARTICLE{Ohnaka2025,
       author = {{Ohnaka}, K. and {Wong}, K.~T. and {Weigelt}, G. and {Hofmann}, K.-H.},
        title = "{High-angular-resolution ALMA imaging of the inhomogeneous dynamical atmosphere of the asymptotic giant branch star W Hya: SiO, H$_{2}$O, SO$_{2}$, SO, HCN, AlO, AlOH, TiO, TiO$_{2}$, and OH lines}",
      journal = {\aap},
         year = 2025,
        month = dec,
       volume = {704},
          eid = {A18},
        pages = {A18},
          doi = {10.1051/0004-6361/202554900},
archivePrefix = {arXiv},
       eprint = {2512.03156},
 primaryClass = {astro-ph.SR},
       adsurl = {https://ui.adsabs.harvard.edu/abs/2025A&A...704A..18O}
}

@ARTICLE{Miller2026b,
       author = {{Miller}, David R. and {Caiazzo}, Ilaria and {Heyl}, Jeremy and {Richer}, Harvey B. and {Hollands}, Mark A. and {Tremblay}, Pier-Emmanuel and {El-Badry}, Kareem and {Rodriguez}, Antonio C. and {Vanderbosch}, Zachary P.},
        title = "{The White Dwarf Initial─Final Mass Relation from Open Clusters in Gaia DR3}",
      journal = {\apj},
         year = 2026,
        month = jan,
       volume = {996},
       number = {1},
          eid = {69},
        pages = {69},
          doi = {10.3847/1538-4357/ae18c8},
archivePrefix = {arXiv},
       eprint = {2510.24877},
 primaryClass = {astro-ph.SR},
       adsurl = {https://ui.adsabs.harvard.edu/abs/2026ApJ...996...69M}
}

@ARTICLE{Miller2026a,
       author = {{Miller}, David R. and {Heyl}, Jeremy and {Tremblay}, Pier-Emmanuel},
        title = "{Investigating the Observational Progenitor Mass Gap in the White Dwarf Initial-Final Mass Relation. I. Cluster Census and Characterization of the First White Dwarfs in the Gap}",
      journal = {arXiv e-prints},
         year = 2026,
        month = jul,
          eid = {arXiv:2607.24941},
        pages = {arXiv:2607.24941},
          doi = {10.48550/arXiv.2607.24941},
archivePrefix = {arXiv},
       eprint = {2607.24941},
 primaryClass = {astro-ph.SR},
       adsurl = {https://ui.adsabs.harvard.edu/abs/2026arXiv260724941M}
}

@ARTICLE{Paxton2011,
       author = {{Paxton}, Bill and {Bildsten}, Lars and {Dotter}, Aaron and {Herwig}, Falk and {Lesaffre}, Pierre and {Timmes}, Frank},
        title = "{Modules for Experiments in Stellar Astrophysics (MESA)}",
      journal = {\apjs},
         year = 2011,
        month = jan,
       volume = {192},
       number = {1},
          eid = {3},
        pages = {3},
          doi = {10.1088/0067-0049/192/1/3},
archivePrefix = {arXiv},
       eprint = {1009.1622},
 primaryClass = {astro-ph.SR},
       adsurl = {https://ui.adsabs.harvard.edu/abs/2011ApJS..192....3P}
}

@ARTICLE{Hofner2018,
       author = {{H{\"o}fner}, Susanne and {Olofsson}, Hans},
        title = "{Mass loss of stars on the asymptotic giant branch. Mechanisms, models and measurements}",
      journal = {\aapr},
         year = 2018,
        month = jan,
       volume = {26},
       number = {1},
          eid = {1},
        pages = {1},
          doi = {10.1007/s00159-017-0106-5},
       adsurl = {https://ui.adsabs.harvard.edu/abs/2018A&ARv..26....1H}
}

@ARTICLE{Decin2021,
       author = {{Decin}, Leen},
        title = "{Evolution and Mass Loss of Cool Ageing Stars: a Daedalean Story}",
      journal = {\araa},
         year = 2021,
        month = sep,
       volume = {59},
        pages = {337-389},
          doi = {10.1146/annurev-astro-090120-033712},
archivePrefix = {arXiv},
       eprint = {2011.13472},
 primaryClass = {astro-ph.SR},
       adsurl = {https://ui.adsabs.harvard.edu/abs/2021ARA&A..59..337D}
}

@ARTICLE{Freytag2023,
       author = {{Freytag}, Bernd and {H{\"o}fner}, Susanne},
        title = "{Global 3D radiation-hydrodynamical models of AGB stars with dust-driven winds}",
      journal = {\aap},
         year = 2023,
        month = jan,
       volume = {669},
          eid = {A155},
        pages = {A155},
          doi = {10.1051/0004-6361/202244992},
archivePrefix = {arXiv},
       eprint = {2301.11836},
 primaryClass = {astro-ph.SR},
       adsurl = {https://ui.adsabs.harvard.edu/abs/2023A&A...669A.155F}
}

@ARTICLE{Decin2019,
       author = {{Decin}, L. and {Homan}, W. and {Danilovich}, T. and {de Koter}, A. and {Engels}, D. and {Waters}, L.~B.~F.~M. and {Muller}, S. and {Gielen}, C. and {Garc{\'\i}a-Hern{\'a}ndez}, D.~A. and {Stancliffe}, R.~J. and {Van de Sande}, M. and {Molenberghs}, G. and {Kerschbaum}, F. and {Zijlstra}, A.~A. and {El Mellah}, I.},
        title = "{Reduction of the maximum mass-loss rate of OH/IR stars due to unnoticed binary interaction}",
      journal = {Nature Astronomy},
         year = 2019,
        month = feb,
       volume = {3},
        pages = {408-415},
          doi = {10.1038/s41550-019-0703-5},
archivePrefix = {arXiv},
       eprint = {1902.09259},
 primaryClass = {astro-ph.SR},
       adsurl = {https://ui.adsabs.harvard.edu/abs/2019NatAs...3..408D}
}

@ARTICLE{Heisler1986,
       author = {{Heisler}, J. and {Tremaine}, S.},
        title = "{The influence of the Galactic tidal field on the Oort comet cloud}",
      journal = {\icarus},
         year = 1986,
        month = jan,
       volume = {65},
       number = {1},
        pages = {13-26},
          doi = {10.1016/0019-1035(86)90060-6},
       adsurl = {https://ui.adsabs.harvard.edu/abs/1986Icar...65...13H}
}

@article{Pham2024,
	title = {Polluting white dwarfs with {Oort} cloud comets},
	volume = {530},
	issn = {0035-8711},
	url = {https://doi.org/10.1093/mnras/stae986},
	doi = {10.1093/mnras/stae986},
	number = {3},
	urldate = {2025-11-20},
	journal = {Monthly Notices of the Royal Astronomical Society},
	author = {Pham, Dang and Rein, Hanno},
	month = may,
	year = {2024},
	pages = {2526--2547},
}

@ARTICLE{maercker2024,
       author = {{Maercker}, M. and {De Beck}, E. and {Khouri}, T. and {Vlemmings}, W.~H.~T. and {Gustafsson}, J. and {Olofsson}, H. and {Tafoya}, D. and {Kerschbaum}, F. and {Lindqvist}, M.},
        title = "{Probing the dynamical and kinematical structures of detached shells around AGB stars}",
      journal = {\aap},
         year = 2024,
        month = jul,
       volume = {687},
          eid = {A112},
        pages = {A112},
          doi = {10.1051/0004-6361/202449643},
archivePrefix = {arXiv},
       eprint = {2405.01222},
 primaryClass = {astro-ph.SR},
       adsurl = {https://ui.adsabs.harvard.edu/abs/2024A&A...687A.112M},
}

@ARTICLE{Kalirai2014,
       author = {{Kalirai}, Jason S. and {Marigo}, Paola and {Tremblay}, Pier-Emmanuel},
        title = "{The Core Mass Growth and Stellar Lifetime of Thermally Pulsing Asymptotic Giant Branch Stars}",
      journal = {\apj},
         year = 2014,
        month = feb,
       volume = {782},
       number = {1},
          eid = {17},
        pages = {17},
          doi = {10.1088/0004-637X/782/1/17},
archivePrefix = {arXiv},
       eprint = {1312.4544},
 primaryClass = {astro-ph.SR},
       adsurl = {https://ui.adsabs.harvard.edu/abs/2014ApJ...782...17K}
}

@ARTICLE{Moe2017,
       author = {{Moe}, Maxwell and {Di Stefano}, Rosanne},
        title = "{Mind Your Ps and Qs: The Interrelation between Period (P) and Mass-ratio (Q) Distributions of Binary Stars}",
      journal = {\apjs},
         year = 2017,
        month = jun,
       volume = {230},
       number = {2},
          eid = {15},
        pages = {15},
          doi = {10.3847/1538-4365/aa6fb6},
archivePrefix = {arXiv},
       eprint = {1606.05347},
 primaryClass = {astro-ph.SR},
       adsurl = {https://ui.adsabs.harvard.edu/abs/2017ApJS..230...15M}
}

@ARTICLE{Kroupa2001,
       author = {{Kroupa}, Pavel},
        title = "{On the variation of the initial mass function}",
      journal = {\mnras},
         year = 2001,
        month = apr,
       volume = {322},
       number = {2},
        pages = {231-246},
          doi = {10.1046/j.1365-8711.2001.04022.x},
archivePrefix = {arXiv},
       eprint = {astro-ph/0009005},
 primaryClass = {astro-ph},
       adsurl = {https://ui.adsabs.harvard.edu/abs/2001MNRAS.322..231K}
}

@ARTICLE{Fischer1992,
       author = {{Fischer}, Debra A. and {Marcy}, Geoffrey W.},
        title = "{Multiplicity among M Dwarfs}",
      journal = {\apj},
         year = 1992,
        month = sep,
       volume = {396},
        pages = {178},
          doi = {10.1086/171708},
       adsurl = {https://ui.adsabs.harvard.edu/abs/1992ApJ...396..178F}
}

@BOOK{Lamers2017,
       author = {{Lamers}, Henny J.~G.~L.~M. and {Levesque}, Emily M.},
        title = "{Understanding Stellar Evolution}",
         year = 2017,
          doi = {10.1088/978-0-7503-1278-3},
          publisher={IOP Publishing},
       adsurl = {https://ui.adsabs.harvard.edu/abs/2017use..book.....L}
}

@ARTICLE{Hwang2022,
       author = {{Hwang}, Hsiang-Chih and {Ting}, Yuan-Sen and {Zakamska}, Nadia L.},
        title = "{The eccentricity distribution of wide binaries and their individual measurements}",
      journal = {\mnras},
         year = 2022,
        month = may,
       volume = {512},
       number = {3},
        pages = {3383-3399},
          doi = {10.1093/mnras/stac675},
archivePrefix = {arXiv},
       eprint = {2111.01789},
 primaryClass = {astro-ph.SR},
       adsurl = {https://ui.adsabs.harvard.edu/abs/2022MNRAS.512.3383H}
}

@ARTICLE{Rein2015,
       author = {{Rein}, Hanno and {Spiegel}, David S.},
        title = "{IAS15: a fast, adaptive, high-order integrator for gravitational dynamics, accurate to machine precision over a billion orbits}",
      journal = {\mnras},
         year = 2015,
        month = jan,
       volume = {446},
       number = {2},
        pages = {1424-1437},
          doi = {10.1093/mnras/stu2164},
archivePrefix = {arXiv},
       eprint = {1409.4779},
 primaryClass = {astro-ph.EP},
       adsurl = {https://ui.adsabs.harvard.edu/abs/2015MNRAS.446.1424R}
}

@ARTICLE{Pham2024a,
       author = {{Pham}, Dang and {Rein}, Hanno and {Spiegel}, David S.},
        title = "{A new timestep criterion for N-body simulations}",
      journal = {The Open Journal of Astrophysics},
         year = 2024,
        month = jan,
       volume = {7},
          eid = {1},
        pages = {1},
          doi = {10.21105/astro.2401.02849},
archivePrefix = {arXiv},
       eprint = {2401.02849},
 primaryClass = {astro-ph.EP},
       adsurl = {https://ui.adsabs.harvard.edu/abs/2024OJAp....7E...1P}
}

@ARTICLE{Rein2012,
       author = {{Rein}, H. and {Liu}, S.-F.},
        title = "{REBOUND: an open-source multi-purpose N-body code for collisional dynamics}",
      journal = {\aap},
         year = 2012,
        month = jan,
       volume = {537},
          eid = {A128},
        pages = {A128},
          doi = {10.1051/0004-6361/201118085},
archivePrefix = {arXiv},
       eprint = {1110.4876},
 primaryClass = {astro-ph.EP},
       adsurl = {https://ui.adsabs.harvard.edu/abs/2012A&A...537A.128R}
}

@ARTICLE{Marigo2007,
       author = {{Marigo}, P. and {Girardi}, L.},
        title = "{Evolution of asymptotic giant branch stars. I. Updated synthetic TP-AGB models and their basic calibration}",
      journal = {\aap},
         year = 2007,
        month = jul,
       volume = {469},
       number = {1},
        pages = {239-263},
          doi = {10.1051/0004-6361:20066772},
archivePrefix = {arXiv},
       eprint = {astro-ph/0703139},
 primaryClass = {astro-ph},
       adsurl = {https://ui.adsabs.harvard.edu/abs/2007A&A...469..239M}
}

@ARTICLE{Zink2020,
       author = {{Zink}, Jon K. and {Batygin}, Konstantin and {Adams}, Fred C.},
        title = "{The Great Inequality and the Dynamical Disintegration of the Outer Solar System}",
      journal = {\aj},
         year = 2020,
        month = nov,
       volume = {160},
       number = {5},
          eid = {232},
        pages = {232},
          doi = {10.3847/1538-3881/abb8de},
archivePrefix = {arXiv},
       eprint = {2009.07296},
 primaryClass = {astro-ph.EP},
       adsurl = {https://ui.adsabs.harvard.edu/abs/2020AJ....160..232Z}
}

@ARTICLE{Brown2022,
       author = {{Brown}, Garett and {Rein}, Hanno},
        title = "{On the long-term stability of the Solar system in the presence of weak perturbations from stellar flybys}",
      journal = {\mnras},
         year = 2022,
        month = oct,
       volume = {515},
       number = {4},
        pages = {5942-5950},
          doi = {10.1093/mnras/stac1763},
archivePrefix = {arXiv},
       eprint = {2206.14240},
 primaryClass = {astro-ph.EP},
       adsurl = {https://ui.adsabs.harvard.edu/abs/2022MNRAS.515.5942B}
}

@ARTICLE{Hut1981,
       author = {{Hut}, P.},
        title = "{Tidal evolution in close binary systems.}",
      journal = {\aap},
         year = 1981,
        month = jun,
       volume = {99},
        pages = {126-140},
       adsurl = {https://ui.adsabs.harvard.edu/abs/1981A&A....99..126H}
}

@ARTICLE{Verbunt1995,
       author = {{Verbunt}, F. and {Phinney}, E.~S.},
        title = "{Tidal circularization and the eccentricity of binaries containing giant stars.}",
      journal = {\aap},
         year = 1995,
        month = apr,
       volume = {296},
        pages = {709},
       adsurl = {https://ui.adsabs.harvard.edu/abs/1995A&A...296..709V}
}

@ARTICLE{Vick2020,
       author = {{Vick}, Michelle and {Lai}, Dong},
        title = "{Tidal evolution of eccentric binaries driven by convective turbulent viscosity}",
      journal = {\mnras},
         year = 2020,
        month = aug,
       volume = {496},
       number = {3},
        pages = {3767-3780},
          doi = {10.1093/mnras/staa1784},
archivePrefix = {arXiv},
       eprint = {1912.04892},
 primaryClass = {astro-ph.SR},
       adsurl = {https://ui.adsabs.harvard.edu/abs/2020MNRAS.496.3767V}
}

@BOOK{Silverman1986,
       author = {{Silverman}, B.~W.},
        title = "{Density estimation for statistics and data analysis}",
         year = 1986,
       adsurl = {https://ui.adsabs.harvard.edu/abs/1986desd.book.....S},
      publisher={Chapman and Hall},
}

@ARTICLE{Hurley2002,
       author = {{Hurley}, Jarrod R. and {Tout}, Christopher A. and {Pols}, Onno R.},
        title = "{Evolution of binary stars and the effect of tides on binary populations}",
      journal = {\mnras},
         year = 2002,
        month = feb,
       volume = {329},
       number = {4},
        pages = {897-928},
          doi = {10.1046/j.1365-8711.2002.05038.x},
archivePrefix = {arXiv},
       eprint = {astro-ph/0201220},
 primaryClass = {astro-ph},
       adsurl = {https://ui.adsabs.harvard.edu/abs/2002MNRAS.329..897H}
}

@ARTICLE{Tokovinin1998,
       author = {{Tokovinin}, A.~A.},
        title = "{On the distribution of orbital eccentricities for wide visual binary stars}",
      journal = {Astronomy Letters},
         year = 1998,
        month = mar,
       volume = {24},
       number = {2},
        pages = {178-179},
       adsurl = {https://ui.adsabs.harvard.edu/abs/1998AstL...24..178T}
}

@ARTICLE{Tokovinin2016,
       author = {{Tokovinin}, A. and {Kiyaeva}, O.},
        title = "{Eccentricity distribution of wide binaries}",
      journal = {\mnras},
         year = 2016,
        month = feb,
       volume = {456},
       number = {2},
        pages = {2070-2079},
          doi = {10.1093/mnras/stv2825},
archivePrefix = {arXiv},
       eprint = {1512.00278},
 primaryClass = {astro-ph.SR},
       adsurl = {https://ui.adsabs.harvard.edu/abs/2016MNRAS.456.2070T}
}

@ARTICLE{Grondin2026,
       author = {{Grondin}, Steffani M. and {Drout}, Maria R. and {Nordhaus}, Jason and {Muirhead}, Philip S. and {Filer}, Bailey and {Laroche}, Alexander and {Webb}, Jeremy J. and {Broekgaarden}, Floor S. and {Chornock}, Ryan and {Kremer}, Kyle and {LeBaron}, Natalie and {Margutti}, Raffaella and {Noughani}, Nikki and {Sears}, Huei and {Tremblay}, Pier-Emmanuel},
        title = "{A Framework for Linking Pre- and Post-Common Envelope Binary Properties with Star Clusters: The First Demonstration with a Massive White Dwarf+M Dwarf Binary in Alessi 12}",
      journal = {arXiv e-prints},
         year = 2026,
        month = jul,
          eid = {arXiv:2607.20611},
        pages = {arXiv:2607.20611},
          doi = {10.48550/arXiv.2607.20611},
archivePrefix = {arXiv},
       eprint = {2607.20611},
 primaryClass = {astro-ph.SR},
       adsurl = {https://ui.adsabs.harvard.edu/abs/2026arXiv260720611G}
}

@ARTICLE{Grondin2024,
       author = {{Grondin}, Steffani M. and {Drout}, Maria R. and {Nordhaus}, Jason and {Muirhead}, Philip S. and {Speagle}, Joshua S. and {Chornock}, Ryan},
        title = "{The First Catalog of Candidate White Dwarf─Main-sequence Binaries in Open Star Clusters: A New Window into Common Envelope Evolution}",
      journal = {\apj},
         year = 2024,
        month = nov,
       volume = {976},
       number = {1},
          eid = {102},
        pages = {102},
          doi = {10.3847/1538-4357/ad7500},
archivePrefix = {arXiv},
       eprint = {2407.04775},
 primaryClass = {astro-ph.SR},
       adsurl = {https://ui.adsabs.harvard.edu/abs/2024ApJ...976..102G}
}

@ARTICLE{AbdelSattar2018,
       author = {{Abdel-Sattar}, Walid and {Mawad}, Ramy and {Moussas}, Xenophon},
        title = "{Study of solar flares' latitudinal distribution during the solar period 2002-2017: GOES and RHESSI data comparison}",
      journal = {Advances in Space Research},
         year = 2018,
        month = nov,
       volume = {62},
       number = {9},
        pages = {2701-2707},
          doi = {10.1016/j.asr.2018.07.024},
       adsurl = {https://ui.adsabs.harvard.edu/abs/2018AdSpR..62.2701A}
}

@ARTICLE{stcyr1999,
       author = {{St. Cyr}, O.~C. and {Plunkett}, S.~P. and {Michels}, D.~J. and {Paswaters}, S.~E. and {Koomen}, M.~J. and {Simnett}, G.~M. and {Thompson}, B.~J. and {Gurman}, J.~B. and {Schwenn}, R. and {Webb}, D.~F. and {Hildner}, E. and {Lamy}, P.~L.},
        title = "{Properties of coronal mass ejections: SOHO LASCO observations from January 1996 to June 1998}",
      journal = {\jgr},
         year = 2000,
        month = jan,
       volume = {105},
       number = {A8},
        pages = {18169-18186},
          doi = {10.1029/1999JA000381},
       adsurl = {https://ui.adsabs.harvard.edu/abs/2000JGR...10518169S}
}

@ARTICLE{ElBadry2021,
       author = {{El-Badry}, Kareem and {Rix}, Hans-Walter and {Heintz}, Tyler M.},
        title = "{A million binaries from Gaia eDR3: sample selection and validation of Gaia parallax uncertainties}",
      journal = {\mnras},
         year = 2021,
        month = sep,
       volume = {506},
       number = {2},
        pages = {2269-2295},
          doi = {10.1093/mnras/stab323},
archivePrefix = {arXiv},
       eprint = {2101.05282},
 primaryClass = {astro-ph.SR},
       adsurl = {https://ui.adsabs.harvard.edu/abs/2021MNRAS.506.2269E}
}

@ARTICLE{Poon2026,
       author = {{Poon}, Michael and {Pham}, Dang and {Bryan}, Marta L. and {Rein}, Hanno and {Dong}, Jiayin},
        title = "{Stellar Separation Shapes Spin─Orbit Alignment in Visual Binaries}",
      journal = {\apjl},
         year = 2026,
        month = may,
       volume = {1002},
       number = {2},
          eid = {L36},
        pages = {L36},
          doi = {10.3847/2041-8213/ae61af},
archivePrefix = {arXiv},
       eprint = {2604.18921},
 primaryClass = {astro-ph.SR},
       adsurl = {https://ui.adsabs.harvard.edu/abs/2026ApJ..1002L..36P}
}

@ARTICLE{Poon2025,
       author = {{Poon}, Michael and {Bryan}, Marta L. and {Rein}, Hanno and {Dong}, Jiayin and {Speagle}, Joshua S. and {Pham}, Dang},
        title = "{Early Evidence for Isotropic Planetary Obliquities in Young Super-Jupiter Systems}",
      journal = {\apjl},
         year = 2025,
        month = dec,
       volume = {994},
       number = {2},
          eid = {L48},
        pages = {L48},
          doi = {10.3847/2041-8213/ae1f0e},
archivePrefix = {arXiv},
       eprint = {2511.04091},
 primaryClass = {astro-ph.EP},
       adsurl = {https://ui.adsabs.harvard.edu/abs/2025ApJ...994L..48P}
}
\bibliographystyle{mnras}

\appendix

\section{Separation Distribution Fitting Procedure}\label{appendix:sep}

In this appendix, we describe the separation data fitting procedure, which closely follows that in \citet{ElBadry2018}.

\subsection{Incompleteness}

For a binary with angular separation $\theta$ and $G$-band magnitude difference $\Delta G$, \citet{ElBadry2018} modeled the companion detection probability as
\begin{equation}
    f(\theta, \Delta G) = \frac{1}{1+(\theta/\theta_0)^{-\beta}}.
\end{equation}
The parameters were empirically constrained to be $\beta\approx10$ and $\theta_0$ described by
\begin{equation}
    \theta_0 (\Delta G) \approx \begin{cases} 
      2.25\text{ arcsec} & ,\Delta G \le 1.5 \text{ mag} \\
      0.9(\Delta G + 1) \text{ arcsec}  & ,\text { otherwise}.
\end{cases}
\end{equation}
Furthermore, the on-sky angular separation can be related to the projected, on-sky separation $s_\mathrm{obs}$ as $\theta = s_\mathrm{obs}/D_i$ where $D_i$ is the distance to the binary system (from Earth).
Thus, we can write the detection probability as $f(s_\mathrm{obs};D_i,\Delta G_i)$, where $D_i$ and $\Delta G_i$ are measured for each binary in the catalog.

\subsection{Likelihood Function}\label{sec:lfuncs}

Using the detection probability $f(s_\mathrm{obs};D_i,\Delta G_i)$, we construct the likelihood that the observed separation data is described by our binary population synthesis model (BPS).
Our likelihood function follows that of \citet{ElBadry2018}.
Let $\phi_{\rm BPS}(s_\mathrm{obs}|\vec{m})$ denote the probability density of projected separation predicted by the model BPS with parameters $\vec{m} = (\nks, d, v_\mathrm{rms})$.
The three parameters correspond to the number of kicks, the magnitude of each kick, and the dimensionality of the kick(s), respectively.
For a sample of $N_\mathrm{obs}$ observed binaries with projected separations $s_{\mathrm{obs},j}$, the likelihood can be written as
\begin{align}
    \mathcal{L}
                &\propto\prod_{j=1}^{N_\mathrm{obs}} \frac{\phi_{\rm BPS}(s_{\mathrm{obs},j}|\vec{m})}{\int_{s_\mathrm{min}}^{s_\mathrm{max}} \dd s'~\phi_{\rm BPS}(s'|\vec{m}) ~f(s'; D_j, \Delta G_j)}.
\end{align}
The quantities $s_\mathrm{min}$ and $s_\mathrm{max}$ are the minimum and maximum separations of $s_{\mathrm{obs}}$, respectively.
In this work, $s_\mathrm{min}=10^{1.5}~\mathrm{au}$ and $s_\mathrm{max}=10^{4.5}\mathrm{~au}$.
For additional details on quality cuts, biases, completeness, and the likelihood function, we refer the reader to \citet{ElBadry2018}.

We now describe how the probability density function $\phi_\mathrm{BPS}$ is found from the binary population synthesis model.
When all binaries have finished evolving for a given set of simulation parameters, we extract the bound systems, record their intrinsic separation, and calculate the projected separation for an observer.
We estimate $\phi_\mathrm{BPS}$ from the projected separations using Gaussian kernel density estimation \citep[e.g.,][]{Silverman1986}.

Finally, we maximize the likelihood function to fit our binary population synthesis model parameters to observed data.
Although in principle, all three parameters can be fitted simultaneously, models with different $N_\mathrm{kicks}$ and $d$ can all reproduce the observed separation distribution well.
Therefore, we fix the kick dimension and number of kicks and find the optimal kick strength for each combination.

\subsection{Mock Observations}\label{sec:mock}

The methodology above allows us to fit our model to an incomplete observational sample.
We also generate mock observations to compare the simulated and observed distributions directly.

To do so, we assign each simulated binary a mock distance $D$ and a $G$-band magnitude difference $\Delta G$.
We draw these quantities by inverse transform sampling of their observed distributions in the catalog of \citet{ElBadry2018}.
It is important to note that this procedure \textit{assumes} that $D$ and $\Delta G$ are independent of each other and of other variables in the system.
This may be the most tenuous assumption we make, however, the resulting mock distributions nonetheless agree reasonably well with the observations.

With the projected on-sky separation, $D$, and $\Delta G$ from the simulated samples, we can calculate the completeness correction function $f(s_\mathrm{obs}, D, \Delta G)$ for each system.
We then use rejection sampling with acceptance probability $f(s_\mathrm{obs},D,\Delta G)$ to construct the mock observed sample.
Specifically, for each system we draw $x\sim\mathrm{Uniform}[0,1]$ and retain the system if $x<f(s_\mathrm{obs},D,\Delta G)$.

\section{Inferring the Eccentricity Power-law Index}\label{appendix:ecc}

\begin{figure}
    \centering
    \includegraphics[width=1.\linewidth]{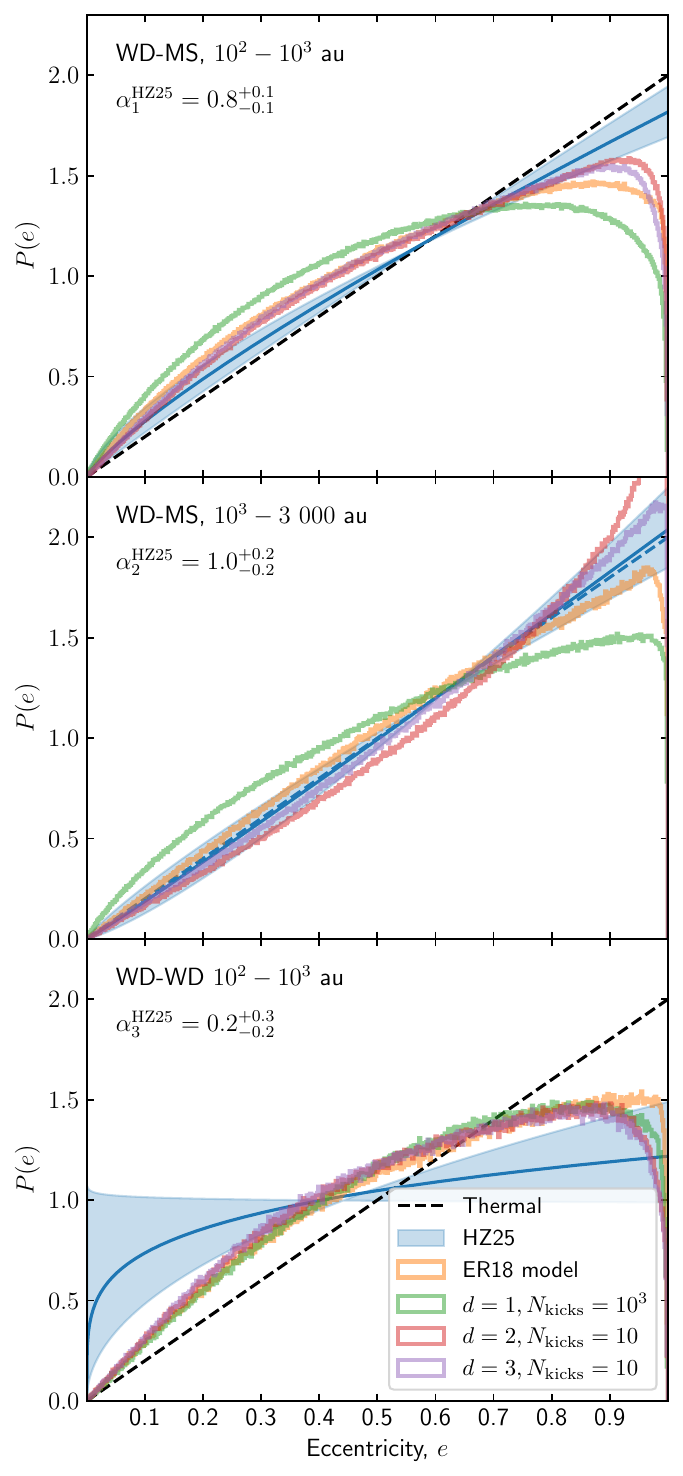}
    \caption{Eccentricity distributions for selected models. Panels correspond to the binary types and projected-separation bins indicated. The adiabatic mass-loss and kick model by \citet{ElBadry2018} is abbreviated as ER18. Shaded regions show the power-law distributions implied by the eccentricity-index constraints of \citet{hwang2025}, abbreviated as HZ25. Colored curves show the simulation predictions, and the dashed line shows the thermal distribution.  The deficit at high eccentricity reflects preferential removal by the common envelope criterion, motivating fits with several upper eccentricity cutoffs, $e_\mathrm{max}$.}
    \label{fig:eobs}
\end{figure}

Here, we describe how to find the best-fit eccentricity power-law index, $\alpha$, from simulation data.
For each $(N_\mathrm{kicks},d)$, we simulate with the best-fitted kick strength found previously, then extract the eccentricities, $e$, and projected separations of the surviving simulated WD--MS and WD--WD binaries.
We estimate $\alpha$ by maximum likelihood following \citet{hwang2025}.
We use the truncated power-law probability density
\begin{equation}
    P(e;\alpha,e_\mathrm{max}) =
    \paren{\frac{1+\alpha}{e_\mathrm{max}^{\alpha+1}}} e^\alpha,
\end{equation}
for $0\leq e < e_\mathrm{max}\leq1$ and $\alpha>-1$.

The maximum eccentricity cutoff $e_\mathrm{max}$ is necessary as the high-$e$ end of the distribution is most susceptible to boundary conditions.
In Fig. \ref{fig:eobs}, we compare the eccentricity distributions for the ER18 adiabatic mass-loss single-kick model, unidirectional $d=1$ kicks with $10^3$ kicks, and $d=2$ and $d=3$ random kicks with 10 kicks, in the same observed separation bins as \citet{hwang2025}.
The figure illustrates why the choice of the upper eccentricity cutoff, $e_\mathrm{max}$, matters.
The deficit of highly eccentric binaries reflects our removal criterion: these systems are particularly susceptible to removal at small pericenter distances.

Then, for a given $e_\mathrm{max}$ cutoff, the likelihood of the retained simulated eccentricities under this model is
\begin{align}
    \mathcal{L} &= \prod_{j=1}^{N_\mathrm{sims}} P(e_j ; \alpha, e_\mathrm{max})\\
    &= (1 + \alpha)^{N_\mathrm{sims}} e_\mathrm{max}^{-N_{\mathrm{sims}}(\alpha+1)} \paren{\prod_{j=1}^{N_\mathrm{sims}}e_j}^\alpha,
\end{align}
where $e_j$ is the eccentricity of the $j$-th retained binary and $N_\mathrm{sims}$ is the number of simulated systems satisfying $0 \leq e_j < e_\mathrm{max}$.
Setting $\partial\ln\mathcal{L}/\partial\alpha=0$ gives the maximum-likelihood estimate, giving the best-fit $\alpha$ for a particular set of simulated eccentricities
\begin{equation}
    \alpha(e_\mathrm{max}) = -\paren{N_\mathrm{sims} \Bigg/ \sum_{j=1}^{N_\mathrm{sims}}\ln \paren{\frac{e_j}{e_\mathrm{max}}}} - 1.
\end{equation}

We perform the procedure above to calculate the best-fit $\alpha$ for binaries within the separation bins in \citet{hwang2025}.
We compute the $\chi^2$ to compare simulated eccentricity indices with observational constraints.
Furthermore, we report the reduced $\chi_\mathrm{reduced}^2=\chi^2/\nu$, using $\nu=3$ degrees of freedom for the three observational constraints.
For each best-fit kick model, no parameters are fitted to these eccentricity constraints, so we subtract no fitted parameters when assigning $\nu$.

\bsp	
\label{lastpage}
\end{document}